\documentstyle[epsf,eqsecnum,floats,preprint,aps,epsfig]{revtex}

\input epsf
\tighten
\def\bx{{\bf x}}
\def\ve{{\varepsilon}}

\def\cE{{\cal E}}

\def\high{\vphantom{\Biggl(}\displaystyle}

\def\mpl{M_{\rm Pl}}
\def\half{\frac{1}{2}}
\def\gev{{~{\rm GeV}}}

\begin{document}
\newcommand{\bm}[1]{\mbox{\boldmath{$#1$}}}
\newcommand{\be}{\begin{equation}}
\newcommand{\ee}{\end{equation}}
\newcommand{\bea}{\begin{eqnarray}}
\newcommand{\eea}{\end{eqnarray}}
\newcommand{\barr}{\begin{array}}
\newcommand{\earr}{\end{array}}

\rightline{}
\rightline{hep-th/0501184}
\rightline{UFIFT-HEP-05-3}
\vskip 1cm

\begin{center}
\ \\
\large{{\bf Inflation from Warped Space }} 
\ \\
\ \\
\ \\
\normalsize{ Xingang Chen }
%%\footnote{\tt email address}
\ \\
\ \\
\small{\em Institute for Fundamental Theory \\ Department of Physics,
University of Florida, Gainesville, FL 32611 }

\end{center}

\begin{abstract}
A long period of inflation can be triggered when the inflaton is held
up on the top of
a steep potential by the infrared end of a warped space. 
We first
study the field theory description of such a model. We then embed it
in the flux stabilized string compactification. Some special
effects in the throat reheating process by relativistic branes are
discussed. We put all these ingredients into a multi-throat brane
inflationary scenario. The resulting cosmic string tension and a
multi-throat slow-roll model are also discussed.
\end{abstract}

\setcounter{page}{0}
\thispagestyle{empty}
\maketitle

\eject

\vfill

\baselineskip=18pt

\section{Introduction}
Inflation\cite{Guth:1980zm,Linde:1981mu,Albrecht:1982wi} provides a
natural mechanism for creating the homogeneity
and flatness of our observable universe. It also gives an elegant way
of generating the
perturbations\cite{Mukhanov:1981xt,Starobinsky:1982ee,Hawking:1982cz,Guth:1982ec,Bardeen:1983qw,Mukhanov:1982nu,Mukhanov:1990me}
which
seed the structure formation. In order for the inflation to
last sufficiently long and then successfully exit to reheat
the universe, the inflaton has to be held up on a potential for a
sufficiently long time. Such a mechanism is achieved most commonly by
a potential
which is very flat on the top. The required flatness is summarized
by the slow-roll conditions. A central problem in inflation has been
to find
a natural realization of such a flat potential in a fundamental
theory. Many years of research in supergravity and string theory
indicates that, while such flat potentials may arise in many
occasions,
they are not generic. Therefore it is important to ask if
inflation can happen given a steep potential, while generating a scale
invariant spectrum. In this paper we study
such a model by making use of the warped space.

Recently warped space has shown its importance in both field
and string theory. It has been proposed as one
of the few possible explanations to the hierarchy
problem\cite{Randall:1999ee}. In string
theory, such space arise as a consequence of flux
compactification\cite{Klebanov:2000hb,Giddings:2001yu},
and play important roles in stabilizing the extra dimensions and
constructing the dS
space and inflationary
models\cite{Kachru:2003aw,Silverstein:2004id,Kachru:2003sx,Silverstein:2003hf,Hsu:2003cy,Firouzjahi:2003zy,Pilo:2004mg,Burgess:2004kv,DeWolfe:2004qx,Iizuka:2004ct,Alishahiha:2004eh,Berg:2004ek,Blanco-Pillado:2004ns,Buchel:2004qg,Chen:2004gc,Berg:2004sj,Shandera:2004zy,Firouzjahi:2005dh,Becker:2005sg}.

In this paper, we study an interesting use of the warped
space in the
context of the inflation. Since the infrared (IR) end of a warped
space generally has a small warp factor, to a bulk
observer at the ultraviolet (UV) end of this warped space,
anything trapped in the IR side moves very slowly in the
extra dimension.
This is because
the speed of light traveling in the extra dimension is
small in the IR end.
In particular this applies to a D3-brane. To a four-dimensional
observer, the extra dimension is the internal space and the position
of the D3-brane in the extra dimension is a
scalar field. Therefore
this provides a new mechanism for the scalar 
field to move very
slowly\cite{Silverstein:2003hf,Alishahiha:2004eh,Chen:2004gc}.

In terms of the gravitationally coupled scalar field theory, 
we will be interested in a scalar field with a relativistic kinetic
term and
rolling down from the top of a steep
potential in a warped internal space. 
The causality restricts the scalar to roll
slowly and we will show that it is quite robust against the steepness
of the potential. The
appearance of the resulting inflation is kind of similar to the
slow-roll inflation: the inflaton stays at the top of the potential
for a long time before it falls down through fast rolling. However the
more detailed nature of these two scenarios are different: for
example, in
slow-roll inflation, the potential is flat and the inflaton is
non-relativistic, while here the potential is steep and the inflaton
is ultra-relativistic during inflation.
This mechanism is
especially interesting in situations where the warped space become
necessary and generic, but the flat potentials are not.
We will call this type of
inflationary models as DBI inflation.

It is important to realize effective field theories of any
inflation model in a unified fundamental theory such as string
theory. In this paper we are interested in the idea of the brane
inflation\cite{Dvali:1998pa,Burgess:2001fx,Dvali:2001fw,Alexander:2001ks,Shiu:2001sy,Quevedo:2002xw}
in the flux stabilized string
compactification\cite{Giddings:2001yu,Kachru:2003aw,Kachru:2003sx}.
The setup is
the orientifold compactification of type IIB string theory. Besides
stabilizing the complex and dilaton moduli, the NS-NS and R-R fluxes
also induce warped space (throats) around various conifold
singularities. Such
warped space carry D3-charges. They attract anti-D3-branes in the bulk
and then annihilate them. Because the D3-charge of a throat is
discrete in multiples of some integers, D3-branes will generally be
created at the IR end of the throat after this
annihilation\cite{Kachru:2002gs,DeWolfe:2004qx}.

We will be particularly interested in those throats with large flux
numbers. The flux-antibrane annihilation process in such a throat
proceeds through quantum
tunneling\cite{Kachru:2002gs,DeWolfe:2004qx}. In the four-dimensional
spacetime
point of view, the annihilation creates a bubble in the false
vacuum. Generally the interior of the bubble is still in a false
vacuum, because there may be a moduli potential for the resulting
D3-branes, or anti-D3-branes in other places of the compact manifold
waiting to be annihilated. If such
a moduli potential is flat enough, the slow-roll inflation can happen
within the bubble.  Under a steep repulsive moduli
potential, normally these D3-branes will quickly roll out and our
universe cannot live in such a bubble. However since here we
have a situation where the D3-branes are trapped in an IR warped
space, such a rolling is subject to a causality constraint, namely
the DBI inflation can happen.

A multi-throat brane inflation model\cite{Chen:2004gc} will be studied
in more details. In
this model, branes generated as above roll out of the brane (B)
throat, triggering the DBI inflation. They reheat and settle
down in the Standard-Model (S) throat. We show that such a model can
generate the right density perturbations with a direct reheating and a
Randall-Sundrum (RS) warp factor. Subtleties of the relativistic brane
reheating\cite{Chen:2004hu}, and its
effect on the density perturbations and on the large flux number are
studied. Other possible cases, for example adding an antibrane (A)
throat, and a multi-throat slow-roll model are also discussed.

The multi-throat configuration provides a unique opportunity to observe
signals of string theory. It gives a hierarchical range of
scales. For low string scale such as the RS setup, we may have chance
to observe strings in colliders. For throats with high string scale,
brane inflation can create cosmic
strings\cite{Jones:2002cv,Sarangi:2002yt,Polchinski:2004ia}. They may
give
observable signals in addition to the density perturbations and the
spectral
index\cite{Jones:2003da,Pogosian:2003mz,Dvali:2003zj,Copeland:2003bj,Leblond:2004uc,Jackson:2004zg,Kibble:2004hq,Damour:2004kw}.
We will discuss the corresponding string tension in
our cases. We also discuss another way strings are produced during the
dS epoch,
which are more general but with less tension.

This paper is organized as follows. In Sec.~\ref{SecField}, we
describe the effective field theory of the DBI inflation. This
includes the zero-mode inflation and the density perturbations. In
Sec.~\ref{SecString}, we embed the field theory in the setup of the
flux compactification, where the inflaton dynamics is described by the
DBI action of D3-branes in warped extra dimensions. Various
constraints coming from the validity of the DBI action and an
interesting stringy suppression mechanism on density perturbations are
discussed. In Sec.~\ref{SecReheating}, we turn to the
reheating process. We emphasize two important processes that arise
quite often for the throat reheating in a multi-throat setup --
the relativistic brane collision and cosmological rescaling. We put
all these ingredients in a multi-throat model in Sec.~\ref{SecMulti}.
In Appendix \ref{SecRollingS}, we discuss branes rolling into a
throat, which is used in Sec.~\ref{SecReheating}, and briefly review
another DBI inflation model. Appendix \ref{SecMultiSR} studies
how the DBI inflation and slow-roll inflation are jointed in
case of a flat potential. This leads to a multi-throat slow-roll
inflation model.
Cosmic strings produced in different cases are discussed accordingly in
Sec.~\ref{SecMulti} and Appendix \ref{SecMultiSR}.

\section{Field theory of DBI inflation}
\label{SecField}
Although the DBI inflation scenario is motivated by string
theory,
the field theory description of the main process during the
inflationary
period can be extracted out independently, and is interesting in its
own right. This describes a scalar rolling
down from a steep potential in a warped internal space. 
In this section, we
will study how inflation arises in this setup and the resulting
density
perturbations. When appropriate, we will mention its connections to
the string model that will be discussed later.
A similar type of model was studied in
Ref.~\cite{Silverstein:2003hf,Alishahiha:2004eh}, where an important
difference is to start the inflaton from
the UV side. The resulting
inflationary scenario has some qualitative differences and will be
compared in the end of Appendix \ref{SecRollingS}.

For later convenience, we will denote the scalar field as $r$, which
is related to the usual scalar field $\phi$ through $r(t,\bx) \equiv
\phi(t,\bx)/\sqrt{T_3}$ (the constant $T_3$ is the
brane tension). The scalar moves in an internal warped space with a
characteristic length scale $R$
\bea
ds^2 = h^2(r) g_{\mu \nu} dx^\mu dx^\nu + h^{-2}(r)
dr^2 ~, 
~~~ h(r)=r/R ~.
\label{WpMetric}
\eea
Here $ds_4^2 = g_{\mu \nu} dx^\mu dx^\nu$ is the four-dimensional
space-time metric. It is highly warped near $r\sim 0$. The scalar
field $r(t, {\bf x})$ can be thought of as a 4-d
hypersurface embedded in 5-d space (\ref{WpMetric}).

The action which governs the gravitationally coupled scalar is given
by
\bea
S = \frac{\mpl^2}{2} \int d^4x \sqrt{-g} R
- T_3 \int d^4x \sqrt{-g} \left[ h^4 \sqrt{1+h^{-4} g^{\mu\nu}
\partial_\mu r \partial_\nu r } - h^4 +V(r) \right] ~.
\label{ActionSG}
\eea
Note that in this paper $V$ has been made dimensionless by
pulling out a factor of $T_3$.
The kinetic term in (\ref{ActionSG}) may be understood as a
generalization of the kinetic term for a
homogeneous scalar in flat four-dimensional space-time
\bea
-T_3 \int d^4 x \left[ h^4 \sqrt{ 1-h^{-4}~ \dot r^2 } \right] ~,
\eea
whose integrand is proportional to the proper length of a relativistic
particle traveling in the warped space. Another familiar limit is the
non-relativistic limit where $|h^{-4} g^{\mu\nu}
\partial_\mu r \partial_\nu r| \ll 1$. The action then reduces to the
minimal case
\bea
-T_3 \int d^4x \sqrt{-g} \left[ \half g^{\mu\nu}
\partial_\mu r \partial_\nu r + V(r) \right] ~.
\label{non-rel-action}
\eea
As we will see, in terms of a D3-brane moving in
extra dimensions, the action (\ref{ActionSG}) comes from the DBI and
Chern-Simons action describing the low-energy effective world-volume
field theory
of a probe brane in the AdS and R-R fields background.

We assume that the potential $V(r)$ has a maximum at $r=0$ and falls
as $r>0$. For a
generic non-flat potential, in the familiar case of
(\ref{non-rel-action}), the scalar will undergo a fast-roll and make
the inflation impossible. Here the highly warped space near $r\sim 0$
plays an important role. The idea is that the scalar velocity is
restricted by the speed of light in the internal space $\dot r \leq
h^2$. Therefore the requirement of slow rolling translates into
the requirement of a small warp factor. This is interesting since an
exponentially large warping is not difficult to find. In fact, it
turns out that there are more stringent constraints coming from, for
example, the strength
of the background which supports the warp factor against the inflaton
back-reaction, and the infrared closed string creation of the dS
back-reaction. These constraints will be discussed in
Sec.~\ref{SecInf} and Sec.~\ref{SecDBI}.

\subsection{The inflation}
\label{SecInf}
We first study the zero-mode dynamics of the scalar inflaton, which
drives the inflation. We ignore the spatial
inhomogeneities of the scalar field so that it is only a function of
time $r(t)$. The four-dimensional metric $g_{\mu\nu}$ is taken to be
\bea
{\rm diag}(-1,a^2(t),a^2(t),a^2(t)) ~,
\label{Metric0}
\eea
where $a(t)$ is
the scale factor. The action (\ref{ActionSG}) then becomes
\bea
S = \frac{\mpl^2}{2} \int d^4x \sqrt{-g} R - T_3 \int d^4x~a^3(t)
\left[ h^4 \sqrt{1- h^{-4} \dot r^2} - h^4 + V(r) \right] ~.
\label{Action}
\eea
The corresponding equations of motion are
\bea
3\left( \frac{\dot a}{a} \right)^2 = \frac{T_3}{\mpl^2} \left(
\frac{h^6}{\sqrt{h^4-\dot r^2}} - h^4 + V \right) ~,~~~~~~~~~~~~~~ 
\label{Eoma} \\
\frac{d}{dt}\left( \frac{h^2 \dot r}{\sqrt{h^4-\dot r^2}} \right) +
3H \frac{h^2 \dot r}{\sqrt{h^4-\dot r^2}} + \frac{2h\partial_r h
(2h^4-\dot r^2)}{\sqrt{h^4-\dot r^2}} - \partial_r \left(h^4 -
V(r)\right) = 0 ~.
\label{Eomr}
\eea
We will only need the information of the
potential near $r \sim 0$ and
expand $V(r)$ as
\bea
V(r) \approx V_i - \half m^2 r^2 ~.
\label{Pot}
\eea
We will start the scalar
inflaton deep in the warped space from $r_0 \sim 0$. A realization of
such an initial condition will be provided in Sec.~\ref{SecString}.

For clarity, we make two approximations to be verified
in the end of this subsection. First, we approximate that during
inflation the potential $V(r)$
stays as a constant $V(0)\equiv V_i$. As we will see, this is because
the inflaton moves over only a very short distance during the
inflation. Second, the kinetic energy of the scalar field,
namely the first two terms on the right hand side of the
Eq.~(\ref{Eoma}), is negligible comparing to the potential $V$. This
is because these two terms are red-shifted by the warped factor
$h^4$. Both assumptions hold because during inflation
the inflaton is held inside the IR region for a sufficiently long
time. This will translate into a not-very-restrictive upper bound on
$V_i$. The Eq.~(\ref{Eoma}) is then significantly
simplified. It
gives a dS space with a Hubble constant
\bea
H = \frac{\dot a}{a} = \frac{\sqrt{V_i T_3}}{\sqrt{3}\mpl} ~.
\eea
From now on we will denote $m^2 \equiv \beta H^2$ as long as $H$ is
a constant.

In the non-relativistic limit $\dot r \ll h^2$, the equation of motion
(\ref{Eomr}) for $r$ reduces to the familiar form
\bea
\ddot r + 3H\dot r + \partial_r V(r) = 0 ~.
\label{Nonrel}
\eea
If $\beta \ll 1$, the potential (\ref{Pot}) satisfies the slow-roll
conditions and the Eq.~(\ref{Nonrel}) determines the slow-roll
velocity. It is also interesting to see how the warp factor will
affect such dynamics and we study it in Appendix
\ref{SecMultiSR}. Here we concentrate on the
more general situation where $\beta \gtrsim 1$. In this case, the
inflaton will be accelerated quickly to become relativistic if
$h^2$ is small enough. 

We thus expand the inflaton evolution around the speed of light
\bea
r = -\frac{R^2}{t} \left(1- \frac{\lambda}{(-t)^p} + \cdots \right) ~,
\label{Expandr}
\eea
where we have chosen the time to run from $-\infty$.
The leading contributions in Eq.~(\ref{Eomr}) come from the second
term
\bea
\frac{3H R^2}{\sqrt{ 2(p-1)\lambda} ~(-t)^{2-\frac{p}{2}} }
\label{Second}
\eea
and the potential term 
\bea
\frac{\beta H^2 R^2}{t} ~.
\label{Potterm}
\eea
The subleading terms are suppressed at least by a factor of $1/Ht$ and
neglected if
\bea
t \ll -H^{-1} ~~~{\rm or}~~ r\ll R^2 H ~.
\label{tUpper}
\eea
The parameters $p$ and $\lambda$ in (\ref{Expandr}) are determined by
matching (\ref{Second}) and (\ref{Potterm}). We get
\bea
r = -\frac{R^2}{t} + \frac{9 R^2}{2\beta^2 H^2} \frac{1}{t^3} + \cdots ~,
\label{rAsymp}
\eea
where the condition
\bea
t \ll - \beta^{-1} H^{-1} ~~~{\rm or}~~ r\ll \beta R^2 H
\label{tUpper2}
\eea
is required for such an expansion.
For the case that we are mostly interested in, $\beta \gtrsim 1$,
the condition (\ref{tUpper}) is stronger than (\ref{tUpper2}). 

As emphasized in Ref.~\cite{Silverstein:2003hf,Chen:2004hu}, the
back-reaction of the relativistic
inflaton can have significant impact on the DBI action. The
condition that such a back-reaction can be neglected can be estimated
as follows. The warping scale caused by the energy density of the
inflaton
field in the internal space is characterized by $R'^4 \sim
\gamma/T_3$,
where $\gamma = h^2/\sqrt{h^4 -\dot r^2}$ is the Lorentz contraction
factor. This scale has to be much smaller than that of the
background $R^4$. Or equivalently, as we will discuss in
Sec.~\ref{SecString}, the
background warped space with $R^4 \sim N/T_3$ can be thought of as
being created by $N$ source D3-branes. The energy density of the
relativistic probe D3-brane should be much smaller than the source for
the back-reaction to be ignored. Using (\ref{rAsymp})
this condition, $\gamma \ll N$, becomes
\bea
t\gg -3N \beta^{-1} H^{-1} ~~~{\rm or}~~ r \gg \frac{\beta R^2 H}{3N}
~.
\label{tLower}
\eea

Let us now summarize the dynamics of the inflaton inside the throat. 
Starting from the place $\high{ r \gg
\frac{\beta R^2 H}{3N}}$  where the back-reaction can be ignored, the
inflaton travels ultra-relativistically
toward the UV side of the warped space under the acceleration of the
potential (\ref{Pot}). The coordinate
velocity is bound by the causality constraint and is very
small. During this period, the inflaton is
held up at the top of the potential for a sufficiently long time to
trigger the inflation.
The Lorentz contraction factor of the inflaton decreases in this
process. Around $r \sim R^2H$, the
inflaton starts to become non-relativistic due to the
increased warp factor. But the coordinate velocity is in fact much
larger. Inflation is ended and the inflaton undergoes a fast-roll
down to the bottom of the potential.
During the whole inflationary period, the inflaton is
relativistic. This period lasts for $\Delta t \sim N \beta^{-1}
H^{-1}$, so the
total number of inflationary e-folds is
\bea
N_{tot} \approx H\Delta t \sim N \beta^{-1} ~.
\label{Nt}
\eea
To have a large $N$, we need $R$ to be bigger than
$T_3^{-1/4}$. For example, if $R \sim 10 T_3^{-1/4}$, we have $N\sim
10^4$. In terms of string theory flux compactification, such a
value is not difficult to find.
In fact, as we will see from a more detailed model in
Sec.~\ref{SecMulti}, a sufficient amount of inflation proceeds even if
$\beta$ is considerably larger than one.
Within the range (\ref{tUpper}) and (\ref{tLower}), the inflaton
position $r$ is related to the latest e-folds $N_e$ by
\bea
N_e \approx H R^2/r \approx \frac{\sqrt{T_3} R}{\sqrt{3}\mpl}
\sqrt{V_i}~ h^{-1} ~.
\label{Ne}
\eea
This expression can be turned around and viewed as a requirement for
$h$ in order to have $N_e$ e-folds of inflation. This is easy to
satisfy since the warp factor is usually exponentially
small. Therefore we will
consider the constraint from the back-reaction (\ref{Nt}) to be
stronger.

\medskip
We have a few comments here.

Besides the lower bound (\ref{tLower}) coming from the back-reaction,
we will also have corrections related to the initial starting point
$r_0$ at $-t_0$, if we assume the inflaton starts there with zero
velocity. This gives the asymptotic behavior (\ref{rAsymp}) a
correction
of order $R^2/t_0$.\footnote{If $-t_0 \ll - \beta^2 N_e^3/H$, 
or $r_0 \ll R^2H/\beta^2 N_e^3$,
this correction does not affect the first two leading terms in
(\ref{rAsymp}), and therefore does not change our analyses. If
$- \beta^2 N_e^3/H<-t_0 \ll -N_e/H$, or $R^2H/\beta^2 N_e^3 < r_0 \ll
R^2H/N_e$, the second term 
in (\ref{rAsymp}) is affected. But this will only lower the velocity
and make the back-reaction smaller. Having larger $-t_0$ will then
decrease the total number of e-folds.} Nonetheless as mentioned,
because the main constraint from the back-reaction on the
total number of inflationary e-folds is usually much stronger than the
requirement of having a relatively large warping, we will always
assume that the
inflaton starts from a small enough $r$
and the abovementioned correction can be ignored. 

The motion of the inflaton within the region where the back-reaction
cannot be ignored is under less precise control so far. A qualitative
description is discussed in\cite{Chen:2004hu}. The time
scale is expected to be roughly of order $N\beta^{-1}H^{-1}$ if we
think of this region as having
an effective warp factor similar to the lower bound
(\ref{tLower}). This will make the inflationary period last even
longer. Since the total number of e-folds (\ref{Nt}) is
already very large, in this paper we assume
this period to be outside of the observable universe.

More importantly, there are other more stringent constraints coming
from the
back-reaction of multiple D3-branes and infrared closed string
creation. We will describe these in terms of strings and
branes in Sec.~\ref{SecDBI}.

Interestingly, the DBI inflation persists even if $\beta
\lesssim 1$. What happens is that, as we decrease $\beta$, a growing
period of slow-roll inflation smoothly matches on to the end of a
long period of DBI inflation. We will study this in
Appendix \ref{SecMultiSR}.
\medskip

We now check the consistency requirement for
the two approximations made in the beginning of this
subsection. First,
the distance $\Delta r \sim R^2 H$ that the
inflaton moves over during the inflation lowers the potential by
$\beta H^4 R^4$. Second,
in Eq.~(\ref{Eoma}), the kinetic energy is proportional to $h^4 \gamma
< R^4H^4\gamma \sim \beta N_e R^4 H^4$. Both are much less than $V_i$
if 
\bea
V_i T_3 \ll \frac{\mpl^4}{\beta N_e R^4 T_3} ~,
\eea
which is very easy to satisfy and normally having $HR \ll 1$ is
enough.

\subsection{Density perturbations}
\label{SecDenPer}
In the previous subsection we have studied the zero-mode evolution of
the
inflaton field and gravitational background. In this subsection we
will study perturbations around it.
In Ref.~\cite{Garriga:1999vw}, Garriga and
Mukhanov have developed a general
formalism to calculate the density perturbations for their kinetic
energy
driven inflation model\cite{Armendariz-Picon:1999rj}. Their analyses
are very general and we can directly 
adapt them here as well. A similar application can be found
in\cite{Alishahiha:2004eh}.

\medskip
Before we start the rigorous derivation, we would like to present
an intuitive approach\cite{Chen:2004gc} which gives a more explicit
interpretation of
the underlying physics in our case. As we have seen, a special
property of the
inflaton in our case is that it travels relativistically and the
corresponding Lorentz contraction factor $\gamma$ is decreasing. If we
choose
at each moment an instantaneous frame which moves at the same speed as
the inflaton, the zero mode velocity of the inflaton vanishes to
this observer. (It is a good approximation for large $N_e$. This is
because $\Delta \gamma/\gamma \approx \Delta N_e/N_e$, so the relative
change in $\gamma$ is negligible in a duration of several e-folds.)
Because of the time dilation, the Hubble constant
is increased by a factor of $\gamma$ to this moving
observer.
We can then use the result of the scalar fluctuations
in the minimally coupled (non-relativistic) case, namely 
$\delta r' \equiv \delta
\phi'/\sqrt{T_3} \approx H\gamma/(2\pi \sqrt{T_3})$. This amplitude is
essentially determined by applying the uncertainty principle to the
inflaton momentum generated within a Hubble horizon of size
$H^{-1} \gamma^{-1}$. After they are stretched outside of the Hubble
horizon, their amplitudes get frozen because they are no longer in
causal contact. We then switch to the lab observer, the horizon size
remains the same since it is in the direction orthogonal to the
velocity. But the frozen scalar amplitude will be reduced by a factor
of $\gamma^{-1}$ because of the relativistic Lorentz contraction.
So we get $\delta r \approx H/(2\pi \sqrt{T_3})$ which is the same as
the slow-roll case, except that the horizon size is now reduced by a
factor of $\gamma^{-1}$. This horizon is also called the sound
horizon.

Because of these scalar inhomogeneities, different spatial part of the
universe will end the inflation at different
time\cite{Hawking:1982cz,Guth:1982ec} (in a gauge where we
set the unperturbed slice synchronous). For small
perturbations $\delta r \ll r$, the time difference is
\bea
\delta t \approx \frac{\delta r_*}{\dot r_*} 
\approx \frac{H}{2\pi h_*^2 \sqrt{T_3}} 
\approx \frac{N_e^2}{2\pi \sqrt{T_3} R^2 H} ~.
\label{Deltat}
\eea
In the third step, Eq.~(\ref{Ne}) is used.
The subscript $*$ means that the variable is evaluated at the time of
the horizon crossing when the corresponding mode is frozen. 
This time delay seeds the large scale structure
formation\cite{Hawking:1982cz,Guth:1982ec,Peebles:1994xt,Liddle:2000cg}.
On the
scale of Cosmic Microwave Background (CMB), the resulting density
perturbation is given by
\bea
\delta_H = \frac{2}{5} \ve_r H \delta t \approx 
\ve_r \frac{N_e^2}{5\pi \sqrt{T_3}R^2} ~.
\label{deltaH}
\eea
In the simplest
case $\ve_r =1$. But for later purpose, we define $\ve_r
\equiv H_r \delta t_r/H \delta t$. Notice here that
we have denoted the Hubble constant $H_r$ during the
reheating
differently from the Hubble constant $H$ during the inflation. In
the usual field theory we normally assume that they are approximately
equal. But applying to the multi-throat string compactification, they
may be very different because the 
reheating can happen in a throat not responsible for the
inflation. Independent warp factors cause the subtlety of a possible
period of cosmological rescaling process in such a throat. This cannot
be described by an effective single scalar field theory and may be
imposed as a boundary condition.
It also has the effect of shifting the wave-number
and rescaling the $\delta t \rightarrow \delta t_r$ by a related
factor. We leave these details specific to
string models to Sec.~\ref{SecReheating}~\&~\ref{SecMulti}.

\medskip
Let us now start to apply the formalism from\cite{Garriga:1999vw}. The
fluctuations
around the zero-mode evolution (\ref{Metric0}) and (\ref{rAsymp}) can
be parameterized in the following way\cite{Mukhanov:1990me}
\bea
ds_4^2 \equiv g_{\mu\nu} dx^{\mu}dx^{\nu} &=&
-(1+2\Phi) dt^2 + (1-2\Phi) a^2(t) d{\bf x}^2 ~, \nonumber \\
r(t,\bx) &=& r_0(t) + \delta r(t,\bx) ~,
\label{Pert}
\eea
where we have added the subscript $0$ to denote the zero-mode
evolution. Following the notation in\cite{Garriga:1999vw}, we denote the
pressure and energy density as
\bea
p(X,r)/T_3 &\equiv& -h^4 \sqrt{1+2h^{-4}X} + h^4 -V ~, \nonumber \\
\ve (X,r)/T_3 &\equiv& \frac{h^4}{\sqrt{1+2h^{-4}X}} - h^4 + V ~,
\label{pande}
\eea
where
\bea
X \equiv \half g^{\mu\nu} \partial_{\mu} r \partial_{\nu}r ~.
\eea
The equations of motion for the perturbations follow from the
Einstein's equations
\bea
\frac{\partial}{\partial t} \frac{\delta r}{\dot r_0} 
&=& \Phi + \frac{2 \mpl^2}{T_3} \frac{c_s^2}{\ve +p}
\frac{\nabla^2}{a^2} \Phi ~,
\label{Eomdeltar}\\
\frac{\partial}{\partial t} (a\Phi) 
&=& \frac{T_3}{2\mpl^2} a(\ve+p) \frac{\delta r}{\dot r_0} ~,
\label{EomPhi}
\eea
where the sound speed $c_s$ is defined as
\bea
c_s^2 \equiv 1+2h^{-4}X ~.
\eea
In (\ref{Eomdeltar}) and (\ref{EomPhi}), the $\ve$, $p$ and $c_s$ are
all evaluated by the zero-mode solutions, which are
\bea
c_s &=& \frac{\sqrt{h_0^4 - \dot r_0^2}}{h_0^2} = \gamma^{-1}
~. \nonumber \\
\ve/T_3 &=& \gamma h_0^4 - h_0^4 + V(r_0) ~, 
\label{epcEvaluate} \\
p/T_3 &=& -\gamma^{-1} h_0^4 + h_0^4 - V(r_0) ~, \nonumber
\eea
Using the new variables $\xi$ and $\zeta$,
\bea
\xi &\equiv& \frac{2\mpl^2 \Phi a}{T_3 H} ~,
\label{xi} \\
\zeta &\equiv& H \frac{\delta r}{\dot r_0} + \Phi ~,
\label{zeta}
\eea
and the relation $\dot H = -\frac{T_3}{2\mpl^2} (\ve + p)$, we can
rewrite the equations of motion as
\bea
\dot \xi &=& \frac{a(\ve+p)}{H^2} \zeta ~, 
\label{Eomxi} \\
\dot \zeta &=& \frac{c_s^2 H^2}{a^3(\ve+p)} \nabla^2 \xi ~.
\label{Eomzeta}
\eea
Further defining
\bea
z &\equiv& \frac{a(\ve+p)^{1/2}}{c_s H} T_3^{1/2} ~, 
\label{z} \\
v &\equiv& z \zeta ~, \nonumber
\eea
we can simplify Eqs.~(\ref{Eomxi}) and (\ref{Eomzeta}) as
\bea
v'' - c_s^2 ~\nabla^2 v - \frac{z''}{z} v = 0 ~,
\label{Eomv}
\eea
where the prime denotes the derivative with respect to the conformal
time $\eta$ defined by $d\eta = dt/a(t)$. Another equation is of first
order and becomes auxiliary.

To evaluate $z''/z$ we use (\ref{z}) and (\ref{epcEvaluate}). The
leading contribution comes from the scale factor $a$ which has the
strongest
time dependence. The next order comes
from $\ve$, $p$ and $c_s$, which all vary more slowly. The
time-dependence of $H$ is neglected. So we get
\bea
\frac{z''}{z} \approx 2 a^2 H^2 \left( 1 + \frac{1}{2N_e} + {\cal
O}\left( \frac{1}{N_e^2} \right) \right)~.
\eea
Hence for large $N_e$, Eq.~(\ref{Eomv}) reduces to the familiar
equation that we encounter in the slow-roll inflation, except for the
presence
of the sound speed $c_s$. As usual, the solution can be obtained by
matching the short wavelength behavior to the long wavelength behavior
at the
horizon crossing. For $k/a \gg H/c_s$,\footnote{The variation of $c_s$
has to be small enough, $c_s'/c_s \ll k c_s/2\pi$. This is satisfied
since $c_s'/c_s = a\dot c_s/c_s \approx aH/N_e \ll c_s k/N_e$.}
the quantum fluctuations of $v$ reduce to those in the flat
space-time,\footnote{The Bunch-Davies vacuum is chosen here. For
discussions on possible
deviations from it, see e.g.~\cite{Schalm:2004xg} and references
therein.}
\bea
v_k \approx \sqrt{\frac{1}{2c_s k}} ~e^{-i c_s k \eta} ~.
\label{vkshort}
\eea
For $k/a \ll H/c_s$, it is also easy to get the solution
\bea
v_k \approx i\frac{H_*^2}{\sqrt{2} k^{3/2} h_*^2 \sqrt{T_3}}
~z ~,
\label{vklong}
\eea
where the coefficient of $z$ is obtained (up to a constant phase) by
matching it to
(\ref{vkshort}) at the horizon crossing $c_{s*} k = a_*H_*$ and using
$z\approx a h^2 c_s^{-3/2} H^{-1} \sqrt{T_3}$. Hence we see the
well-known phenomenon that, in terms of $\zeta$,
the quantum fluctuations (\ref{vkshort}) evolves to the frozen
classical perturbations (\ref{vklong}).
We also see that the horizon size is ${c_s}_* H_*^{-1} =
\gamma_*^{-1} H_*^{-1}$, agree with the previous intuitive
argument.
Under the assumption of instant and efficient reheating, the
perturbations of the scalar field is transformed into density
perturbations. The corresponding spectral density is
\bea
{\cal P}_{\cal R}(k) \equiv \frac{1}{2\pi^2} k^3 |\zeta_k|^2 
= \frac{H_*^4}{4 \pi^2 h_*^4 T_3} ~,
\eea
where $\zeta_k$ is the Fourier mode of $\zeta$ defined in
(\ref{zeta}).
The density perturbation $\delta_H$ is related to the spectral density
${\cal P}_{\cal R}(k)$ by
\bea
\delta_H^2 \equiv \frac{4}{25} {\cal P}_{\cal R}(k) ~.
\eea
So we recover the result (\ref{deltaH}) (except for a difference
between $H_r \delta t_r$ and $H \delta t$ which we discuss below). 

To compare with the previously mentioned physical interpretation, we
obtain the relation between $\zeta$ and $\Phi$ using (\ref{EomPhi}) and
(\ref{zeta})
\bea
\zeta &=& \frac{5\ve + 3p}{3(\ve + p)} \Phi + \frac{2\ve}{3(\ve+p)}
\frac{\dot \Phi}{H} \nonumber \\
&\approx& \frac{2 V}{3h^4 \gamma} \left( \Phi + \frac{\dot \Phi}{H}
\right) ~. 
\eea
The second term is smaller than the first by a factor of $1/Ht
\approx -1/N_e$. Since $V \gg h^4 \gamma$, we have $\zeta \gg
\Phi$. This means that the first term in (\ref{zeta}) dominates. The
physical interpretation of this term fits in our
previous intuitive
arguments in the convenient gauge choice. So a possible jump of the
Hubble constant from $H$ to $H_r$ and the time delay from $\delta t$
to $\delta t_r$, imposed as an approximate boundary
condition at the
reheating, is translated into a jump in $\zeta$ by a factor of
$\ve_r$. So the
density perturbation will have an additional factor $\ve_r$ (as long
as $\ve_r \zeta \gg \Phi$).
Such a mechanism is provided when we discuss more reheating details in
Sec.~\ref{SecReheating} and arises quite generally in some string
models in
Sec.~\ref{SecMulti} and Appendix~\ref{SecMultiSR}.

\section{DBI inflation in string theory}
\label{SecString}
It is important to ask how the field theory described in
the previous section may be embedded in string
theory. One natural place to realize it
is to use the mobile D3-branes in the flux stabilized string
compactification. This was described in a multi-throat brane
inflation scenario\cite{Chen:2004gc}. In this setup, the position of
branes in the extra
dimensions is the inflaton as in the brane
inflation\cite{Dvali:1998pa}, and the warped extra dimensions
corresponds to the warped internal space.

Giddings, Kachru and Polchinski
(GKP)\cite{Giddings:2001yu,Klebanov:2000hb} show that,
near a conifold
singularity in type IIB string compactification on a Calabi-Yau
manifold, the presence of NS-NS
and R-R three-form fluxes on two dual cycles induces the gravitational
and R-R charges similar to those of the transverse D3-branes. The
equivalent D3-charge is
\bea
N=M K ~,
\label{throatcharge}
\eea 
where $K$ and $M$ is the number of
NS-NS and R-R fluxes respectively, and the characteristic length scale
$R$ of the resulting warped space is given by
\bea
R^4 = \frac{27}{4} \pi g_s N
\alpha'^2 ~.
\label{Rvalue}
\eea
In addition, this warped space has a minimum warp factor in
the IR end
\bea
h_{\rm min} \sim \exp(-2\pi K/3M g_s) ~.
\label{hmin}
\eea
The fluxes generally fix the complex moduli and the
axion-dilaton. A non-perturbative superpotential is used to
stabilize the
K\"ahler moduli and antibranes are introduced to lift the vacuum to dS
space\cite{Kachru:2003aw}.\footnote{Alternatives are studied
in\cite{Burgess:2003ic,Saltman:2004sn}.}

A multi-throat configuration is a generalization of such a
setup, which contains many throats of different warp factors in
different places in the extra dimensions. 
It will be interesting to construct it explicitly, but in this paper
we assume its existence.
We add the D3-branes whose moduli
in throats are the
candidate inflatons. The volume stabilization for the extra
dimensions and the interactions between the D3 and D7-branes
will generate potentials for these D3-brane moduli. 
Details of such interactions are quite complicated and still under 
active studies. Specifically in this paper we will be interested
in the following
situation. Consider the situation where the D3-brane moduli receive
quadratic potentials with mass-squared of ${\cal
O}(H^2)$ or larger. 
This is actually a generic situation as we have seen in the
eta-problem of the slow-roll inflation model building.
In order to have slow-roll inflation, these
contributions have to cancel each other to a certain precision. The
tuning involved depends on the mass range of the
contributing terms and adjustable parameters. 
Here we do not address the origin of these mass
contributions. (Studies can be found
in\cite{Kachru:2003sx,Firouzjahi:2003zy,Berg:2004ek,Berg:2004sj,Shandera:2004zy}.)
But we do not require the
abovementioned fine-tuned cancellations. In the multi-throat setup, we
assume some throats have negative
mass-squared, and some have positive mass-squared. We note
that these potentials are repulsive or attractive for the D3-brane
moduli, but not the (fixed) positions of the throats.\footnote{Here
we are only interested in throats located at various extrema of the
D3-brane moduli potential profile. It is
important to see
that in what situation this can arise naturally (for example, for
throats sitting at orbifold fixed points), or tuning has to be
involved.}
The potential that we considered in
(\ref{Pot}) corresponds to those repulsive ones.

An immediate question is then how the D3-branes can start from the IR
end of a repulsive throat. This can be done by considering the
dynamics of anti-D3-branes in this setup. Like D3-branes,
throats will attract and annihilate anti-D3-branes. This process
undergoes through the
flux-antibrane
annihilation\cite{Kachru:2002gs,DeWolfe:2004qx}. However there are two
important
differences between the flux-antibrane annihilation and
brane-antibrane
annihilation. First, if the number $p$ of the anti-D3-branes is much
smaller than the R-R flux number $M$, this annihilation proceeds
through
quantum tunneling. So the anti-D3-branes in these throats can have
different lifetime. This is necessary if antibranes are used to lift
the AdS vacuum and provide the inflationary
energy\cite{Kachru:2003aw,Kachru:2003sx}.
Second, more important to our current
discussion, a number of D3-branes will generally be created in the
flux-antibrane annihilation.
The reason is that when the anti-D3-branes annihilate against
the NS-NS fluxes, the total D3-charge of the throat can only change in
steps of $M$ according to (\ref{throatcharge}). Unless $p$ is a
multiple of $M$, D3-branes will be
generated in the end of the annihilation to conserve the
D3-charge. The moduli of these D3-branes become the inflaton required
in our DBI inflation.

\subsection{DBI action and its validity}
\label{SecDBI}
The low energy world-volume dynamics of a probe D3-brane in a warped
space such as (\ref{WpMetric}) is described by the Dirac-Born-Infeld
(DBI) and Chern-Simons action\cite{Aharony:1999ti} 
\bea
S= T_3 \int d^4 \xi \left[ -\sqrt{ -{\rm det} \left( \partial_a X^M
\partial_b X^N G_{MN} \right) } 
- \frac{1}{4!} \epsilon^{a_1 \cdots a_4}
\partial_{a_1} X^{M_1} \cdots \partial_{a_4} X^{M_4} C_{M_1 \cdots
M_4} \right] ~.
\label{DBICS}
\eea
$T_3$ is the D3-brane tension, and $\xi^a$ $(a=0,1,2,3)$ are the
D3-brane world-volume coordinates. The functions $X^M(\xi^a)$
$(M=0,1,2,3,4)$
describes the embedding of the D3-brane in the ambient space $X^M$,
where $X^\mu \equiv x^\mu$ $(\mu=0,1,2,3)$ and $X^4 \equiv r$. (We
ignore the other angular directions.) We are
interested in branes transverse to the extra dimension $r$. In this
case, the DBI action restricts the longitudinal scale of the brane to
comove
with the warped background\cite{Chen:2004hu}. Hence we can choose the
convenient choice
$x^\mu = \xi^\mu$ $(\mu=0,1,2,3)$ throughout the evolution (which is
no longer true in
Sec.~\ref{SecRescaling}).
We can then
denote the embedding slice as $r(t,x^i)$, which can be regarded as a
scalar field on the
four-dimensional spacetime. This scalar
describes the position and fluctuations of the brane in the transverse
direction.

For D3-brane the R-R four-form
potential is
$C_{0123} = -\sqrt{-g} ~h^4(r)$, where the coupling $\sqrt{-g}$ is
ensured by the four-dimensional Lorentz invariance with the convention
$\epsilon^{0123}\equiv 1$.\footnote{More explicitly, the four-form
potential $C_{\mu_0 \cdots \mu_3} = \epsilon^{\nu_0 \cdots \nu_3}
g_{\nu_0 \mu_0} \cdots g_{\nu_3 \mu_3} \frac{1}{\sqrt{-g}} h^4(r)$.} 
With the 
addition of other possible potentials $V(r)$, the action (\ref{DBICS})
leads to (\ref{ActionSG}). These additional potentials provide the
inflationary energy. They can come from anti-D3-branes sitting inside
the other throats, D3-D7 brane interactions, or related volume
stabilization.

The validity of the DBI action requires that the energy involved in
the effective field theory be much smaller than the mass of the
massive W-bosons stretching between the probe brane and the
horizon\cite{Aharony:1999ti,Silverstein:2003hf}. From (\ref{rAsymp})
this requirement, i.e.~$\dot r/r \ll r/\alpha'$,
becomes $R^2 \gg \alpha'$, which is in the region where we trust the
supergravity background. More importantly, the probe dynamics is
guaranteed
only when the back-reaction of the D3-brane is
small\cite{Silverstein:2003hf,Chen:2004hu}. This is the main
constraint that we used in Sec.~{\ref{SecInf}. Now we can understand
it more easily in this context. The warped space is the same as the
near-horizon region of a stack of $N$ D3-brane source. The
relativistic effect increases the proper energy density of the probe
D3-brane
by a Lorentz contractor factor $\gamma$. If we treat the gravitational
field strength of such a relativistic brane to be increased by roughly
the same amount, we need $\gamma \ll N$ in order to neglect such a
back-reaction.

There are other effects that are more restrictive than the lower bound
(\ref{tLower}). First, the number of D3-branes created by the
flux-antibrane annihilation is $M-p$. If all these branes stick
together and
exit the throat, the back-reaction will be increased by a factor of
$M$ for $p \ll M$. So the total number of inflationary e-folds is
reduced to
\bea
N_{tot} \sim K \beta^{-1} ~,
\label{NtSmall}
\eea
which we approximate as $\sqrt{N} \beta^{-1}$.

Second,
because the string scale is red-shifting towards the IR end,
the Hubble expansion will be able to create closed strings some place
in the throat. This is possible when the proper Hubble energy becomes
comparable to the string scale, i.e.~$h^{-4} H^4 \approx
\alpha'^{-2}$. But in this subsection we are more interested in its
effect on the
background metric which is responsible for the brane speed limit. Such
effect only gets significant when the energy density of the closed
string gas/network becomes comparable to the source. It
will then smear out the background metric and the effective warp
factor
will no longer decrease. Such a critical warp factor $h_c$ can be
estimated by $h_c^{-4} H^4 \sim N T_3$, where the left
hand side is the proper energy density of the closed string
gas/network
smeared out
in $r<r_c$, and the right hand side is the proper energy density of
the source
brane (or the equivalent fluxes). Using $R^4 \sim N/T_3$, we get $r_c
\sim R^2 H/\sqrt{N}$. So
the total number of e-folds is reduced to $\sqrt{N}$.

So, for a stack ($M$) of branes, we will estimate $N_{tot}$ as
$\sqrt{N}\beta^{-1}$ from (\ref{NtSmall}), while for a single brane,
$N_{tot}$ can be estimated as $\sqrt{N}$.

\subsection{Stringy quantum fluctuations on D3-branes}
\label{SecStringyQF}
According to Sec.~\ref{SecDenPer}, the field quantum fluctuations on
the D3-branes generate the density perturbations
\bea
\delta_H = \frac{2}{5} \ve_r H \delta t \approx \ve_r
\frac{2N_e^2}{5\tilde N} ~, ~~~~~ \tilde N \equiv \sqrt{27 n_B N_B/8} ~, 
\label{DPString}
\eea
where we have used (\ref{deltaH}), (\ref{Rvalue}), $T_3 = (2\pi)^{-3}
g_s^{-1} \alpha'^{-2}$ and considered $n_B$ number of mobile
D3-branes. The corresponding spectral index is
\bea
n_s -1 \approx -\frac{4}{N_e} ~, ~~~~ \frac{dn_s}{d\ln k} \approx
-\frac{4}{N_e^2}~,
\label{nsString}
\eea
which is red and running negatively.

Red-shifted string scale also makes possible the open string creation
on the mobile branes 
and modifies the field theory calculations of the density perturbations
at some scale. This can be most easily seen by considering
the following kinematic bound on the brane transverse fluctuations for
the moving observer\cite{Chen:2004gc}. The
quantum
fluctuations are generated within a
Hubble time and then get stretched out of the horizon. For the moving
observer, the Hubble time is $\gamma^{-1} H^{-1}$. The longest
distance that the brane fluctuations can travel in the transverse
direction is then $\gamma^{-1} H^{-1} h^2$, where $h^2$ is the speed
of light. In the field theory calculation (\ref{Deltat}),
the fluctuation amplitude is $\gamma \delta r \approx \gamma h^2
\delta t$, where we have restored the $\gamma$ factor for the moving
observer. It
satisfies the kinematic bound only if 
\bea
\gamma^2 H \delta t \lesssim 1 ~. 
\label{KBound1}
\eea
This bound also has a dynamical interpretation. Using (\ref{Deltat}),
Eq.~(\ref{KBound1}) is translated into
\bea
\gamma H \lesssim \sqrt{2\pi} T_3^{1/4} h ~.
\eea
This roughly means that the Hubble energy of the dS space has to be
smaller than the red-shifted string scale, which is the valid region
for field theories.

We note that the zero-mode field theory analyses should still remain
valid, although the perturbation analyses break down beyond
(\ref{KBound1}). As long as the background can be trusted under the
conditions that we
discussed in Sec.~\ref{SecDBI}, the only fact used for the zero-mode
is the relativistic speed-limit constraint.

We can also rewrite (\ref{KBound1}) in terms of the latest e-folds
$N_e$ using $\gamma
\approx \beta N_e/3$ [from (\ref{rAsymp})] and (\ref{DPString}),
\bea
N_e \lesssim \sqrt{\frac{3}{\beta}} \tilde N^{1/4} ~.
\label{NeBound}
\eea
Hence, comparing to the naive extension of (\ref{DPString}) beyond
(\ref{NeBound}),
the bound (\ref{KBound1}) offers a suppression mechanism for
larger scales. It is
interesting that such a mechanism is built in without adding any extra
features to the model. 

Let us
here simply suppose that for modes beyond (\ref{NeBound}) the bound is
saturated and study some of its
properties. The density perturbation is then
\bea
\delta_H \approx \ve_r \frac{2}{5\gamma^2} 
\approx \ve_r \frac{18}{5 \beta^2 N_e^2} ~.
\label{DPString2}
\eea
The spectral index,
\bea
n_s - 1 \approx \frac{4}{N_e} ~, ~~~~~ 
\frac{dn_s}{d\ln k} \approx \frac{4}{N_e^2} ~,
\label{Boundns}
\eea
is now blue and running positively. Also, (\ref{Boundns}) have to
be smoothly connected to (\ref{nsString}) through a transition
region. Of course
here we only studied the bound, and a full stringy treatment will be
desirable to give more accurate account. Then we will have an
interesting possibility to observe the
stringy effects: branes, coming from an extremely infrared region
(B-throat), imprint stringy information on their world-volume in terms
of quantum fluctuations and bring them to our world (S or A-throat).

\section{Throat reheating by relativistic branes}
\label{SecReheating}
Reheating after inflation is important to populate the
universe. In brane inflation,
this is achieved by brane collision and annihilation in the S (or A)
throat. In our model, this is sometimes caused by
ultra-relativistic branes.
In this section, we discuss two important processes for such a
reheating\cite{Chen:2004hu}, namely the relativistic collision and
the cosmological rescaling.

\subsection{Annihilation versus collision}
\label{SecAvsC}
We first discuss the direct string production in brane annihilation.
The string
dynamics in brane-antibrane annihilation is
described by Sen's boundary conformal field theory of rolling
tachyon\cite{Sen:2002nu,Sen:2002in,Sen:2004nf}.
Ref.~\cite{Lambert:2003zr} has studied the
one-point function on the disk diagram
in this rolling tachyon background
and show that it is capable of releasing all the brane energy to
closed strings. What happens to the D3-anti-D3-branes is that the
initial inhomogeneities on the brane world-volume will grow and
eventually make them disconnected D0-anti-D0-branes, which then emit
all the energy to a non-relativistic coherent state of heavy closed
strings.

Since the Standard Model will have to live on some surviving
D3-branes or anti-D3-branes, open string creation on such residue
branes will be important for the Big Bang. Loop diagrams with one end
on the rolling tachyon and another on the residue
branes\cite{Chen:2003xq} then become
interesting (see Fig.~\ref{decay} (B)). This is because the
exponentially growing oscillator modes\cite{Okuda:2002yd} in
Sen's boundary state
will create virtual closed strings and contribute to the loop
diagrams. Due to their rapid time dependence, these are candidate
competing diagrams against the disk and partially release
brane energy to open strings. However, there are other loop diagrams
with both ends on the rolling tachyon (see Fig.~\ref{decay} (C)). They
only create
closed strings. Although only a limit amount of information is known
on such diagrams, it is not difficult to see that the evolution of (C)
is much faster than (B), since both ends of (C) are time-dependent
while only one end of (B) is\cite{Chen:2003xq}. So again closed
strings
are dominantly produced in this process.\footnote{We assume that the
difference between the closed and open string couplings is not big.}
(It is possible that subsequently the heavy closed strings
decay to both massless closed and open strings. This cosmological
consequence deserves further
studies\cite{Barnaby:2004gg}.)

\begin{figure}[t]
\begin{center}
%\epsfxsize=8cm
%\epsfbox{decay.eps}
\epsfig{file=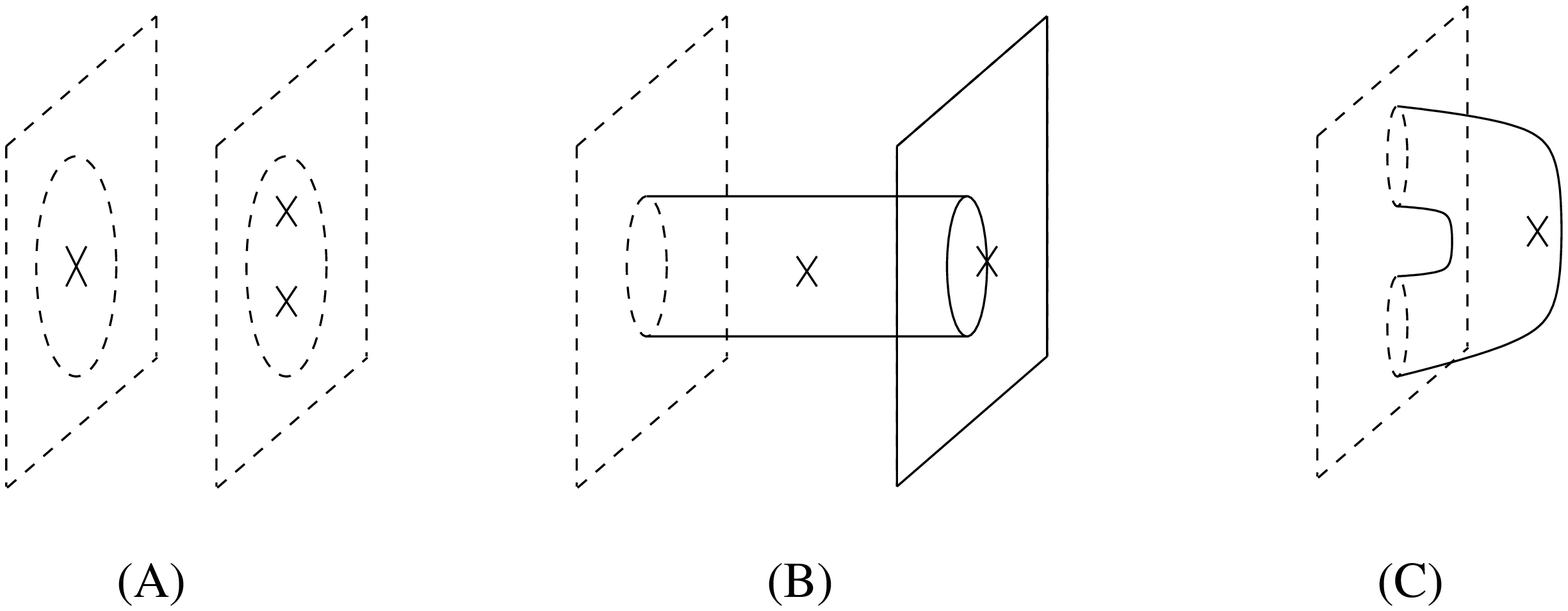, width=12cm}
\end{center} 
\medskip
\caption{Closed and open string creation from brane annihilation. The
branes in dashed lines are decaying ones, and the branes in solid
lines are
the surviving ones. (A) are some
disk diagrams responsible for closed string creation. (B) is a loop
diagram creating closed and/or open strings. The separation between
the branes is in terms of the
world-sheet distance, but branes may not be spatially separated. (C)
is a
loop diagram only creates closed strings.}
\label{decay}
\end{figure}

The annihilation process is important when the colliding branes and
antibranes are non-relativistic. For example in KKLMMT, if we assume
that the slow-roll conditions hold all the way from the UV entrance
to IR end, the brane velocity will remain far below the speed
limit. However in our case, there may not be a direct relation between
the 
inflationary energy scale, which can come from a steep moduli
potential, and the total warping of the S-throat. Hence
the velocity of the D3-branes may be much faster and
there may exist a region in the S-throat where the branes move
relativistically. Such
fast-rolling D3-branes will cause interesting effects on the reheating
details.

The first feature is that the probe branes can become
ultra-relativistic, and the maximum value of its Lorentz contraction
factor is determined by the D3-charge of the background throat. 

To illustrate, let us consider a quadratic attractive potential in the
S-throat
\bea
V_S = \half m_S^2 r^2 ~,
\label{SPot}
\eea
with a positive $m_S^2$ [see (\ref{Smallm})].
Consider $n$ D3-branes rolling out of the B-throat enter this S-throat
directly. The total inflationary potential $V_i$ in (\ref{Pot}) is a
net contribution of the repulsive potential (\ref{Pot}) of the B-throat
and the attractive potential (\ref{SPot}) of the S-throat. After
inflation, this
potential
is converted to the D3-brane kinetic energy when they are in the
S-throat (but still away from the IR end). This provides the initial
velocity for the D3-branes. We denote this velocity as $v_0$ and it is
given by 
\bea
\half n v_0^2 \approx V_i ~.
\label{Initialv}
\eea
A detailed dynamics of such D3-branes can be solved using the DBI
action, and we can find the corresponding place where the probe
back-reaction becomes important. We discuss this in more details in
Appendix.~\ref{SecRollingS}. 
Here let us summarize the relevant main results.

It turns out that as long as the initial velocity $v_0$ satisfies 
\bea
v_0 < \frac{\mpl^2}{T_3 R_S^2 (n N_S)^{1/2}} ~,
\eea
the gravitational coupling of
these probe branes can
be ignored. The resulting dynamics then becomes very simple. It is
determined by the conserved energy density
\bea
\cE/n = \frac{h^6}{\sqrt{h^4-\dot r^2}} - h^4 + V_{\rm net}(r) ~.
\label{ConservedE}
\eea
The D3-branes go through three different phases after the
inflation. In the first stage
they are non-relativistic and accelerated by the potential (\ref{Pot})
and 
(\ref{SPot}) (mainly in
the UV sides of the B and S-throat) to reach a velocity $v_0$. Such a
velocity reaches the speed-limit at $h \approx \sqrt{v_0}$ in the
S-throat and the
branes enter the second relativistic phase. During this phase the
energy density (\ref{ConservedE}) is dominated by the first term,
i.e.~the kinetic energy. The proper spatial volume of the branes
shrinks and
the conserved coordinate energy density is converting from the
brane
tension to the relativistic kinetic energy. [This does not happen in
the non-relativistic phase although the proper spatial volume is also
shrinking because of the cancellation from the R-R field, which is the
second term in (\ref{ConservedE}).] The Lorentz contraction
factor is increasing as $\gamma \approx \half v_0^2 h^{-4}$. At 
\bea
h_r \approx \sqrt{v_0} n^{1/4}/N_S^{1/4} ~,
\label{Rescalingh}
\eea
$n\gamma$ becomes ${\cal O}(N_S)$ and
the probe back-reaction becomes important. The D3-branes then enter a
non-comoving phase.
We will have more to say about this phase in the next subsection.

As long as the reheating happens after the first phase, the energy
transfer is dominated by relativistic collision rather than
annihilation. In terms of direct open string creation,
this process
does not have the abovementioned problem associated with the brane
annihilation. Namely, in
Fig.~\ref{decay}, regarding both branes as colliding ones, the
interaction between the colliding branes is only in terms of diagrams
like (B).
We will estimate the energy density of the created open strings to be
in the same order of magnitude as the collision energy density.
Some interesting properties of the relativistic brane collision are
studied in\cite{McAllister:2004gd}.

\subsection{Cosmological rescaling}
\label{SecRescaling}
We now discuss the second effect closely related in the same
process. If the reheating happens during the second phase discussed
above, the reheating energy density is still approximately the same as
the inflationary energy density, as in the non-relativistic
annihilation case.\footnote{For annihilation this is true if we assume
that the energy transfer to open strings
during the reheating is rapid and 
efficient.} It
is only the way of energy transfer that has been changed from the
annihilation to ultra-relativistic collision (which is good in terms
of direct open string creation). However, this is no
longer true if the reheating happens in the third non-comoving
phase. We will argue that such a phase will introduce effects not
captured in an effective field theory, for example, a jump in the
Hubble constant.

Although the precise mathematical description of the brane
dynamics when back-reaction is significant is unavailable, we can
think of
an analogy of two identical stacks of branes approaching to each
other. Because their energy
density are similar, one will not feel the space being exponentially
warped by
another. Therefore the longitudinal scale of the brane does not
significantly
contract. A similar phenomenon for the relativistic branes should also
happen. Where
this takes place is taken to be at $h_r$ given in (\ref{Rescalingh}),
where the energy density of the
relativistic probe branes is comparable to the source branes (or the
equivalent fluxes). Starting
from $h_r$, the warped background becomes negligible to those probe
branes, and their proper volume is no longer contracting
significantly.

\begin{figure}[t]
\begin{center}
%\epsfxsize=6cm
%\epsfbox{rescaling.eps}
\epsfig{file=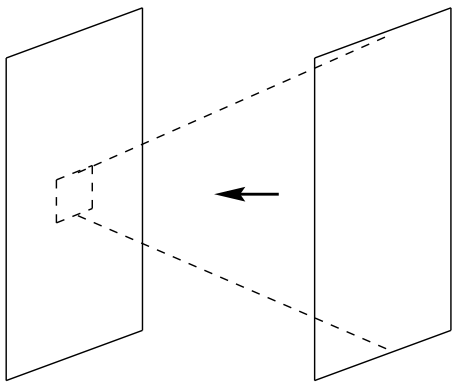, width=6cm}
\end{center} 
\medskip
\caption{The cosmological rescaling for branes in the non-comoving
region where back-reaction is large. The brane in dashed lines in the
IR side
indicates its longitudinal scale if it had followed the DBI action,
which is the scale of the Poincare observer after the background
restoration.}
\label{rescaling}
\end{figure}

Once these branes collide with other branes at the IR end, they will
oscillate and expand. Their energy density is decreasing through
expansion or
radiation. In the mean while, the background is restoring. In fact
it does not take too much expansion to reduce the energy density of
these heated branes, say to one tenth of the original
value. After that, they can again be treated as a probe of the
background.
Since we want this process to be connected to the standard Big Bang,
we will be interested in the Poincare observer on the D3-branes. To
this observer, in the end of the
restoration process, the Planck mass takes the usual value
in the sense of Randall and Sundrum. 
This coordinate choice of such an IR Poincare observer is important,
the scale of such a choice is indicated in Fig.~\ref{rescaling} by a
dashed brane. (The proper energy is independent of such a choice.)
To this observer the space-time
inhomogeneity scale on the
probe D3-branes has changed. This is illustrated in
Fig.~\ref{rescaling}. These inhomogeneities have been geometrically
rescaled by a factor of $g_r = h_r/h_S$, where
$h_S$ is the total warping of the S-throat. In the previous example,
\bea
g_r \sim \frac{n^{1/4} v_0^{1/2}}{N_S^{1/4} h_S} ~.
\label{gr}
\eea

To obtain an order of magnitude estimate, we will ignore the
fast restoration process and simply apply the rescaling factors of
$g_r$ 
to the corresponding length scale, time duration, or energy scale with
respect to their values at $h_r$. 
These effects, not described in a scalar field theory, are then
approximated as imposing effective boundary conditions in the
beginning of the reheating.
For example, the time 
difference $\delta t$ of the inflation ending is geometrically
increased by a factor of $g_r$; the Hubble constant is reduced by a
factor of $g_r^{-2}$ because the energy density is geometrically
decreased by a factor of $g_r^{-4}$. Such rescaling effects can reduce
the $\zeta$ in (\ref{zeta}), and therefore
the density perturbations, as we will see in an example of the
next section.

\section{A multi-throat model}
\label{SecMulti}
A multi-throat brane inflationary scenario has been described
in\cite{Chen:2004gc}, also in the introduction and
Sec.~\ref{SecString}. So
here we only briefly summarize some of the main points. We start
by looking at the anti-D3-branes in the multi-throat
configuration. They are attracted toward the throats, either
annihilate
against the fluxes through classical process, or stay inside in a
quasi-stable state and annihilate through quantum tunneling. The end
products are generally some D3-branes. For those throats (B-throats)
having
potentials like (\ref{Pot}) for the D3-brane moduli, D3-branes will
exit. The DBI inflation discussed in
Sec.~\ref{SecField} \& \ref{SecString} then takes place. 
These branes eventually settle down in throats (S or A) with
attractive D3-brane moduli potential, or in the bulk.
The purpose
of this section is to make this model more quantitative and improve
the calculations by taking into
account the aspects described in
Sec.~\ref{SecStringyQF} \& \ref{SecReheating}. We also discuss the
tension of the cosmic strings created at the end of the inflation from
brane annihilation/relativistic collision, and during the Hagedorn
transition of the dS epoch.

We first consider two-throat case with only B and S-throats, where the
S-throat is defined to have a RS warp factor.
The Hubble constant is simply 
\bea
H = \frac{\sqrt{V_i T_3}}{\sqrt{3}\mpl} ~.
\label{H}
\eea
The inflationary potential can be dominantly provided by the
antibranes in the S-throat, or a moduli potential. This
will be discussed in Case A and Case B, respectively. 

The initial velocity $v_0$ of $n$ number of D3-branes entering the
S-throat is determined 
by the moduli potential. 
In Case A we have
\bea
\half n v_0^2 < V_i \approx 2 n_S h_S^4 ~.
\eea
In Case B we have 
\bea
\half n
v_0^2 \approx V_i ~.
\label{vVCaseB}
\eea
In the latter
case, the velocity does not have
to be as small as in the former, because the height of the moduli
potential is not related
to the S-throat warp factor. Then the rescaling will generally happen
in a deep throat. In
Case C, we consider the addition
of an A-throat, where antibranes there are the main source of
$V_i$.

\medskip
{\bf Case A:}
If the reheating process involving brane collision and/or annihilation
happens before the DBI action breaks down in the S-throat, we have the
usual relation $H \approx H_r$ (assuming an efficient reheating to
open strings). Such a
situation
happens when the initial velocity of the brane is small so that
\bea
h_S > h_r ~,
\label{smallv0}
\eea
where $h_r$ is given in (\ref{Rescalingh}).
For example, if the inflationary energy is dominated by $n_S$
antibranes
at the end of the S-throat, we have $V_i \approx 2 n_S h_S^4$. The
D3-brane kinetic energy density ($\sim \half n v_0^2$) has to
be
smaller than $V_i$ since the moduli potential is not dominant. Hence
the
condition (\ref{smallv0}) is satisfied. In such a case, we notice that
the density
perturbation
\bea
\delta_H \approx 2N_e^2/5\tilde N 
\label{deltaHCaseA}
\eea
is independent of the
warp factor (and $V$), so $h_S$ can take e.g.~$e^{-30}$ to
incorporate the RS model. However at the same time fitting the
observation $\delta_H \approx 1.9 \times 10^{-5}$
requires a very large
$N_B$\cite{Firouzjahi:2005dh,Chen:2004gc} (similar
to\cite{Alishahiha:2004eh}). We take
$N_e \approx 32$, because the inflation is driven by the electro-weak
scale of the S-throat, 
and get $N_B \sim 2.4 \times 10^9$ (estimating $n_B \sim \sqrt{N_B}$).
The total number of inflationary e-folds is $5\times 10^4$ for $\beta
\approx 1$.
If we require the stringy suppression for $\delta_H$ discussed in
Sec.~\ref{SecStringyQF} happens near the largest observable scales,
e.g.~$(N_e)_c 
\approx 29$, we need $\beta \approx 15$. Then the total e-folds
becomes $3 \times 10^3$.

This is our simplest case, but it remains to be seen how
naturally we can get such a large $N_B$. (Getting a large $N_B$
through orbifolding is discussed in\cite{Alishahiha:2004eh}.)
It is interesting to note that $N_B$ can be significantly reduced if
the density perturbations are seeded by (\ref{DPString2}) as we shall
discuss
more in the later comments. In the next, we will discuss the case
where the possible cosmological rescaling helps to reduce the density
perturbations, and therefore $N_B$.

\medskip
{\bf Case B:}
As we emphasized, in our scenario, the inflationary energy does not
have to be correlated with the warp factor of the S-throat. It can
also be
sourced by the steep moduli potential.
Then the reheating Hubble constant $H_r$ is
different from $H$ if the reheating happens in the non-comoving
region in the S-throat. It is determined by the energy density $V_r$
on the reheated branes after the background restoration. This can be
estimated following the description of Sec.~\ref{SecRescaling},
\bea
V_r \approx \alpha \cdot \half n v_0^2 \cdot g_r^{-4} 
\approx \alpha N_S h_S^4 ~. 
\label{Vr}
\eea
In the first step, we approximate the first factor $\alpha
\sim 0.1$, that is, assume that the
D3-branes can 
be treated again as a probe when their energy density is reduced to
one tenth of the source. Reasonable variation of $\alpha$ will not
significantly affect our later estimates. The second factor is the
conserved coordinate energy density of the relativistic branes in the
comoving region. The
last factor is the rescaling factor. The result can also be
simply understood as follows. As long as branes enter the non-comoving
region, the final Lorentz contraction factor
$N_S$ is determined by the strength of the
background and is independent of the initial brane velocity or the
place where
the DBI action breaks down.
The reheating Hubble constant is now
\bea
H_r \approx \frac{\sqrt{\alpha N_S T_3} h_S^2}{\sqrt{3}\mpl} ~.
\label{Hr}
\eea
The reheating time delay after the rescaling process is
\bea
\delta t_r \sim g_r \delta t \approx \frac{g_r N_e^2}{H \tilde N} ~,
~~~~~\tilde N \equiv (27n_B N_B/8)^{1/2} ~,
\eea
where Eqs.~(\ref{DPString}) is used.
The density perturbation can be estimated as
\bea
\delta_H 
&\approx& H_r \delta t_r 
\sim \frac{g_r \alpha^{1/2} N_S^{1/2}
N_e^2 h_S^2}{\tilde N V_i^{1/2}} ~. \\
&\sim& \frac{\alpha^{1/2} N_S^{1/4} h_S}{V_i^{1/4}} 
\frac{N_e^2}{\tilde N} ~.
\label{DenPert}
\eea
In the last step, Eqs.~(\ref{gr}) and (\ref{vVCaseB}) are used. 
The first factor in (\ref{DenPert}) is the effect of the rescaling. At
$h_S \sim h_r$ (and $\alpha \sim 1$), it smoothly goes to one and we
recover (\ref{deltaHCaseA}).

We turn Eq.~(\ref{DenPert}) around and use the measured density
perturbation at the corresponding e-fold to determine the
responsible inflationary potential,
\bea
V_i^{1/4} \sim \frac{\alpha^{1/2} N_S^{1/4} h_S N_e^2}
{\delta_H \tilde N} ~.
\label{Vfromdelta}
\eea

As we discussed in Sec.~\ref{SecStringyQF}, there is a natural
suppression mechanism
for the density perturbations at long wavelengths. This happens at
the critical e-folding
\bea
(N_e)_c \approx \sqrt{3/\beta} {\tilde N}^{1/4} ~.
\label{Nec}
\eea
If this is responsible for the observed CMB suppression near the IR
end,
we can determined $\tilde N$. This is the strategy that we will use in
the following to determine the values of $\tilde N$ and $N_e$.

To do this we first estimate the total number of e-folds needed to
account for the observable universe. We focus on the largest scale
$R_0 \approx 10^{42} \gev^{-1} \approx 10^4 {\rm ~Mpc}$ near the IR end
of the CMB. The
corresponding scale $R_r$ at the time of the
reheating can be estimated by the relation
\bea
R_r \approx \frac{T_0}{T_r} R_0 ~,
\label{Rr}
\eea
where $T_0 \approx 2.7 ~{\rm K}$ is the current temperature and the
reheating temperature $T_r \approx
V_r^{1/4}$.
On the other hand, the Hubble length after rescaling is
\bea
(l_H)_r \equiv g_r l_H \sim g_r \gamma^{-1} 
\frac{\mpl}{\sqrt{V_i T_3}}
~.
\label{lH}
\eea
Here the factor $g_r$ is the rescaling effect discussed in
Sec.~\ref{SecRescaling}. The factor $\gamma^{-1}$, also known as the
sound speed, is the
relativistic effect discussed in Sec.~\ref{SecDenPer}. 
$\gamma$ can be calculated using (\ref{rAsymp}), $\gamma \approx
\beta N_e/3$.

Equations (\ref{Vfromdelta}), (\ref{Rr}) and (\ref{lH}) tell us the e-fold
corresponding to the IR end of the CMB,
\bea
N_e = \ln \frac{R_r}{(l_H)_r} 
\approx 78 + \ln\frac{T_3^{1/4} h_S}{\mpl} + 3 \ln N_e + 
\ln \frac{\beta \alpha^{1/4} N_S^{1/4}}{\tilde N} ~.
\label{NeforCMB}
\eea
$\tilde N$ and $N_e$ can be
determined by
requiring that the $(N_e)_c$ in (\ref{Nec}) is several (e.g.~three)
e-folds below the $N_e$ in
(\ref{NeforCMB}) (setting $\beta \approx 1$ here).
Therefore the inflationary energy scale (\ref{Vfromdelta}) largely
depends on the IR scale $h_S T_3^{1/4}$ of the S-throat.

If we assume that the S-throat solves the hierarchy problem according
to Randall and Sundrum by setting the IR scale $h_S T_3^{1/4}$ to be
around TeV, this 
determines the $N_e$ and $(V_i T_3)^{1/4}$ regardless of the actual
value of $T_3$. We get
\begin{eqnarray}
N_B \sim 8\times 10^6 ~, ~~~ n_B \sim \sqrt{N_B} ~, ~~~
\tilde N \sim 2.8 \times 10^5 ~, \nonumber \\
~~~ (N_e)_c \approx 40 ~, ~~~
N_e \approx 43 ~, ~~~ g_r \sim e^5 ~,  \\
(V_i T_3)^{1/4} \sim 10^6 \gev ~.
\label{ViEx1}
\end{eqnarray}
In this estimation, we used $\alpha \sim 0.1$, $\beta \sim 1$ and $N_S
\sim 10^4$. Among the 48 e-folds of the horizon stretching required to
account for the homogeneity
and flatness of our observable universe, 43 e-folds is given by the
inflation, and the last five e-folds is given by the rescaling.
The total number of inflationary e-folds is $N_{tot} \sim \sqrt{N_B}/\beta
\sim 3\times 10^3$ for $\beta \sim 1$.

\medskip
{\bf Case C:}
There are other possibilities. Let us consider adding an A-throat and
the
inflation ends by brane-antibrane annihilation in this throat.
One option is to assume that the hierarchy problem is not or only
partially solved by the RS mechanism, e.g.~we live in an A-throat.
Then for example in case A (replacing the subscript $S$ with $A$),
$h_A T_3^{1/4}$ needs not be TeV. From Eq.~(\ref{Ne}), the $N_e$
e-folds of inflation happens as long as $h_A$ and $h_B$ satisfy the
relative relation\cite{Chen:2004gc}
\bea
\frac{h_B}{h_A^2} \lesssim \sqrt{\frac{2n_A}{3}}
\frac{R_B \sqrt{T_3}}{\mpl N_e} ~.
\eea
Note that this includes the case where the only throat required is the
B-throat, namely $h_A=1$.
The upper bound on the inflation scale comes from the current
experimental
tensor modes bound\cite{Peiris:2003ff,Tegmark:2003ud}, 
$\high{ {\cal P}^h/{\cal P}_{\cal R} = \frac{8}{75\pi^2}
\frac{V_iT_3}{\delta_H^2 \mpl^4} < 0.5 }$, which gives
\bea
V_i T_3/\mpl^4 < 1.7 \times 10^{-8} ~.
\label{TensorBound}
\eea

Another interesting option is to further add an S-throat. Then the
reheating may
happen either because some branes coming out of the B-throat enter
both the A and
S-throats\cite{Chen:2004gc,Firouzjahi:2005dh}, or the KK modes of the
decay products in the A-throat are
transfered to the S-throat\cite{Barnaby:2004gg}. In either
case, all branes in the A-throat have to be annihilated, and the fate
of closed strings in the A-throat or how much magnitude
of density perturbations can be transfered from A to S deserve
further investigations\cite{Barnaby:2004gg}.

\medskip
We have a few additional comments on various aspects of this model.

{\bf Parameter dependence:} There are some uncertainties in the
estimations: for example in case B, 
the detailed rescaling and background restoration process
parameterized by $\alpha$ in (\ref{Vr}), and the steepness of the
D3-brane moduli potential parameterized by $\beta$. The former only
weakly affects (\ref{Vfromdelta}) and (\ref{NeforCMB}).
The variation of $\beta$ changes the total number of e-folds. More
importantly it changes the $(N_e)_c$ in (\ref{Nec}). The spectral
index (\ref{nsString}) [and (\ref{Boundns})] does not depend
on the overall variation of $\delta_H$ that we talked about.

{\bf Tensor modes:} In our model the condition for the 
inflation to happen is not restricted by the inflationary energy
scale. So the tensor modes bound is very easy to satisfy. For example
in Case B,
$H/\mpl \sim 10^{-25}$; Case C is more flexible. We leave
non-Gaussianity feature for future studies.

{\bf Large ${\bf N_B}$:} For example in
Case B, $N_B$ is $8\times 10^6$. This requires the NS-NS and R-R flux
number to be a few thousands. In GKP compactification, the total D3
charge of all throats and (anti)branes equals  to $\chi/24$, where
$\chi$ is
the Euler number of the corresponding fourfold in F-theory. The
largest value we know of is $\chi/24 =
75852$\cite{Klemm:1996ts,Douglas:2004zg}. It is so
far not clear if we have an actual maximum value, but it seems that
the
model described in this section works if the Euler number is on the
larger side.

We should emphasize here that the large $N_B$ in Case B is no longer
due to the fitting of $\delta_H$. After taking into account of the
non-comoving rescaling, such a degree of freedom goes to the factor
$V_i$ in 
Eq.~(\ref{DenPert}). $N_B$ is large because
we want to calculate most
part of the density perturbations for our observable universe by
the conventional field theory, and only invoke the
open string quantum fluctuations at the early epoch as a
mechanism to suppress the IR end of CMB. The relation (\ref{Nec})
typically gives a
large $N_B$. 

However it remains an open possibility that considerable
part of
the universe is seeded by stringy quantum fluctuations. If this is the
case, the only constraint on $N_B$ is $\sqrt{N_B}/\beta \gtrsim N_e$
for $\sqrt{N_B}$ number of branes, or $\sqrt{N_B} >N_e$ for a single
brane.
Using (\ref{DPString2}), the density perturbations
$\delta_H \approx 1.9 \times 10^{-5}$ is fit around $N_e \approx 55$
by
choosing $\beta \approx 8$ ($\ve_r=1$). We only require $N_B \gtrsim
2\times 10^5$ (or $N_B \gtrsim 3\times 10^3$ for a single brane),
which amounts to a few hundreds (or tens) of flux numbers.

{\bf String statistics:} There is another interesting angle to look at
the large $N_B$ aspect. If the underlying parameters used to
determine the total number of inflationary e-folds is not directly
correlated with
the degrees of freedom used for the string
statistics\cite{Douglas:2004zg}, these
two subjects can be independent of each other. However in our model we
have seen a clear relation between the total
e-folds and flux number, $N_{tot} \sim \sqrt{N_B}$ for $\beta \approx
1$. Such a large factor adds a significant weight to the
statistics and may push the selection rule to favor a compactification
with a large Euler number.

In fact, we expect the same qualitative argument to apply to the
slow-roll case as well, since more fluxes provide more degrees of
freedom to adjust the shape of the potential, although the relation is
much less explicit.

{\bf Cosmic strings:}
The brane inflation gives interesting mechanism to produce cosmic
strings after
inflation\cite{Jones:2002cv,Sarangi:2002yt,Polchinski:2004ia}. The D
or F-strings are produced during
brane-antibrane annihilation through Higgs mechanism or
confinement. Their properties are further studied
in\cite{Jones:2003da,Pogosian:2003mz,Dvali:2003zj,Copeland:2003bj,Leblond:2004uc,Jackson:2004zg,Kibble:2004hq,Damour:2004kw}.
The situation for the relativistic brane collision is
similar, because right after collision the gauge symmetry is also
restored
due to high energy density transfered from the relativistic
branes. In the subsequent cosmological evolution,
this string
network approaches the attractor scaling solution, and can be observed
if their tension is large enough. For strings in the A-throat, the
tension is given by (e.g.~F-strings)
\bea
G \mu_{F} = \sqrt{\frac{g_s}{32\pi}} \frac{T_3^{1/2} h_A^2}{\mpl^2}
~.
\label{Tension1}
\eea

Because strings will be generated if the temperature reaches the
Hagedorn temperature\cite{Englert:1988th,Polchinski:2004ia}, there is
another interesting way that strings
can be produced in any inflationary model in the multi-throat
configuration.
Same as we discussed in the end of Sec.~\ref{SecDBI}, Hubble
energy can exceed the string scale in a deep throat. We take such a
relation to be 
\bea
\frac{H}{2\pi} \gtrsim \frac{h}{4\pi \alpha'^{1/2}} ~,
\eea
where the left hand side is the Gibbons-Hawking temperature and the
right hand
side is the red-shifted Hagedorn temperature.
Therefore any throat with a warp factor 
\bea
h \lesssim \left(\frac{2}{9\pi^3} \right)^{1/4} 
\frac{\sqrt{V_i} T_3^{1/4}}{g_s^{1/4} \mpl}
\label{hHagedorn}
\eea
is filled with strings. After inflation, strings will stay inside each
of these throats
by gravitational attraction and evolve. Hence each throat
independently contributes to the string network.
The corresponding string tension is
\bea
G\mu_F \lesssim \frac{1}{12\pi} \frac{V_i T_3}{\mpl^4} ~.
\label{Tension2}
\eea
This is typically lower than (\ref{Tension1}). 

If observable, we have a very clear signal for the multi-throat
compactification -- an isolated spectrum of the highest string tension
(\ref{Tension1}), whose existence
depends on the existence of the A-throat, followed by a dense spectrum
of lower tension (\ref{Tension2}). The string tension is associated
with the
inflation scale, so in Case A and B they are too weak to be
observed. In Case C, the bound (\ref{TensorBound}) gives
the A-throat string tension
\bea
G\mu_A < 9 \times 10^{-6} \sqrt{g_s/n_A} ~,
\eea
which may be observed by future experiments, 
and the Hagedorn string tension
\bea
G\mu_H < 4 \times 10^{-10} ~.
\label{GH}
\eea
Although the tension is much weaker, the latter can be
enhanced by effects of the multiple throats (e.g.~more signals).

Such a spectrum also arises in a multi-throat slow-roll model that we
will discuss in Appendix \ref{SecMultiSR}. There we will also have a
significant lower bound on $G\mu_A$.

\section{Concluding remarks}
In this paper we have studied a DBI inflation model in both
field and string theory as an alternative to the slow-roll
inflation. The inflation in such a model is achieved by a
inflaton field held on the top of a steep potential by the IR end of a
warped space. 

This model is realized in the multi-throat string compactification
setup. We demonstrate that, at least at the level of orders of
magnitude, this model can
simultaneously produce many interesting features. This
includes the large number of inflationary e-folds; density
perturbations of the right order of magnitude while incorporating the
Randall-Sundrum model with a direct reheating; a scale-invariant
spectrum with interesting features; and a possible 
mechanism for the infrared suppression on CMB.

Many issues remain to be studied in more details. For example, more
detailed information on moduli potential profiles of the multi-throat
GKP compactification with D3-branes; the
stringy quantum fluctuations at the early stage; the detailed
analysis of the non-Gaussian feature (in some cases properly taking
into account the
rescaling or stringy effects); various back-reactions such as the closed
string creation
in the IR B-throat due to the dS back-reaction, and the rescaling
process during the reheating due to the probe brane back-reaction; and
some global aspects of this scenario such as its genericness and
eternity.

\acknowledgments 
I would like to thank Jose Blanco-Pillado, Cliff Burgess, Michael
Douglas, Gia Dvali, Lisa Everett,
Gregory Gabadadze, Daniel Kabat, Louis Leblond, Juan Maldacena,
Massimo Porrati, Zongan 
Qiu, Pierre Ramond, Saswat Sarangi, Sarah Shandera, Gary Shiu, Charles
Thorn, Erick Weinberg, Richard Woodard, and especially Daniel Chung,
Hassan Firouzjahi and Henry Tye for many valuable discussions.
I would also like to thank Eva Silverstein for useful communications
and a suggestion on the terminology.
This work was supported in part by the Department of Energy under
Grant No.~DE-FG02-97ER-41029.

\appendix
\section{Rolling branes in S-throat}
\label{SecRollingS}
In this appendix, we are interested in how $n$ number of D3-branes
roll in the
S-throat according to the DBI action. Most of the
results can be found in
Refs.~\cite{Silverstein:2003hf,Alishahiha:2004eh,Chen:2004hu}. The
main results have
been used in Sec.~\ref{SecReheating}.

This dynamics is described by the equations of motion (\ref{Eoma}) and
(\ref{Eomr}), but the potential $V(r)$ is replaced by
\bea
V(r) = \half m_S^2 r^2 ~.
\label{PotS}
\eea
Here $m_S \gtrsim H$ so gives a generic steep attractive potential.
The region that we will be interested in is well
within the S-throat. 
We start with the initial velocity $-v_0$. Such a velocity is picked up
after the D3-branes
fall down the potentials (\ref{Pot}) and most of (\ref{PotS}). We first check
that this velocity does not change much by the Hubble friction term
if the warp factor $h \gg \sqrt{v_0}$.

The Hubble constant is mainly given by the kinetic energy of the
non-relativistic brane after it falls down the potential, 
\bea
H \approx \frac{\sqrt{n T_3}}{\sqrt{6} \mpl} v_0 ~.
\label{Hv0}
\eea
The velocity change due to the second
Hubble friction term in (\ref{Nonrel}) can be estimated as follows
\bea
\Delta \dot r 
\sim \frac{\sqrt{n T_3}}{\mpl} v_0^2 \Delta t 
< \frac{\sqrt{n T_3} R_S}{\mpl} v_0 ~.
\eea
Since we can estimate $\sqrt{n T_3} R_S/\mpl \sim
N_S^{1/4} n^{1/2} T_3^{1/4}/\mpl$, which is typically not much bigger
than one,
the brane velocity remains in the
same order of magnitude.

Because of the decreasing warp factor, this velocity reaches the
speed-limit around $h \sim \sqrt{v_0}$. We then return to the
Eq.~(\ref{Eomr}). The factor of $H$ in the second term is still given
by the conserved (mostly kinetic) energy (\ref{Hv0}) if this term is
negligible to the first term. This amounts to a comparison between $H$
and
$d/dt \sim 1/t$. If $t \ll \mpl/(\sqrt{n T_3} v_0)$, the gravitational
coupling in (\ref{Eomr}) can indeed be ignored. We have a
conserved coordinate energy density (\ref{ConservedE}) with $\cE
\approx n v_0^2/2$ ($V_{\rm net}$ is approximately zero in the IR side
of the S-throat). The D3-branes
become relativistic
\bea
r = \frac{R_S^2}{t} - \frac{R_S^{10}}{7 v_0^2 t^9} + \cdots ~.
\label{Asympr2}
\eea
Now the time coordinate is chosen to be positive.
If $t \gg
\mpl/(\sqrt{n T_3} v_0)$, we need solve (\ref{Eoma}) and
(\ref{Eomr}), and get
\bea
r &=& \frac{R_S^2}{t} - \frac{\alpha R_S^2}{t^5} + \cdots ~,
\nonumber \\
a &\propto& t^p ~,
\label{Asympr3}
\eea
where
\bea
\alpha &=& \frac{ \left(-1+\sqrt{1+3 n m_S^2 T_3
R_S^4/2\mpl^2}\right)^2}{6 m_S^4} ~, \\
p &=& \frac{n T_3 R_S^4}{2 \mpl^2 \sqrt{6\alpha}} ~.
\eea
Let us first consider the case
\bea
m_S^2 \ll \mpl^2/(n T_3 R_S^4) \sim \mpl^2/(n N_S) ~,
\label{Smallm}
\eea
where these parameters are simplified,
\bea
\alpha \approx \frac{3 (n T_3)^2 R_S^8}{32 \mpl^4} ~, ~~~~~~~
p \approx 2/3 ~.
\eea
In this case the slow expansion of the scale factor $a(t)$ is driven
by the
kinetic energy of the D3-branes. One can check that the second terms
of
(\ref{Asympr2}) and (\ref{Asympr3}) match each other at the turning
point $t \sim \mpl/(\sqrt{n T_3} v_0)$.

We now summarize the D3-brane dynamics following the DBI action.
For $h \gg \sqrt{v_0}$, the brane is moving
non-relativistically. Within $v_0 R_S \sqrt{n T_3}/\mpl \lesssim h
\lesssim \sqrt{v_0}$, the brane travels relativistically. The
gravitation is approximately decoupled in this region, and the brane
dynamics (\ref{Asympr2}) is determined by a conserved coordinate
energy density (\ref{ConservedE}). In
this period, the Lorentz contraction factor increases as $\gamma
\approx \half v_0^2 h^{-4}$. For $h \lesssim v_0 R_S \sqrt{n
T_3}/\mpl$,
the gravitation coupling becomes important. The kinetic energy of the
moving brane
drives the expansion of the scale factor as in (\ref{Asympr3}). The
Lorentz contraction factor increases as $\gamma \approx h^{-2}
\mpl^2/(n T_3 R_S^2)$. 

The probe back-reaction is important if $n\gamma \sim N_S$. For $\high{ 
v_0 < \frac{\mpl^2}{T_3 R_S^2 (n N_S)^{1/2}} }$, 
it happens within the second
interval. This is what we used in Sec.~\ref{SecReheating}.
\medskip

We now consider the large mass-squared
\bea
m_S^2 \gg \mpl^2/(n T_3 R_S^4) \sim
\frac{\mpl^2}{n N_S} ~.
\label{Largem}
\eea
In this
case, Refs.~\cite{Silverstein:2003hf,Alishahiha:2004eh} show that the
inflation is possible. Now the spatial
expansion is driven by the potential energy of the branes. In
(\ref{Asympr3}), we have
\bea
\alpha \approx \frac{n T_3 R_S^4}{4 m_S^2 \mpl^2} ~, ~~~~
p \approx \frac{m_S \sqrt{n T_3 R_S^4}}{\sqrt{6} \mpl} ~.
\eea
The probe back-reaction restricts
\bea
h > \frac{n^{1/4} m_S^{1/2} \mpl^{1/2}}{N_S^{1/2} T_3^{1/4}} ~.
\eea
Therefore in this setup the IR space below this region does not play
important role for inflation because the brane
velocity cannot further decrease. Accordingly, for inflation
to happen, we need a large mass-squared in the moduli potential to
give
a high inflationary energy. This requires (\ref{Largem}) because
the total e-folds is given by
\bea
N_{tot} = p \ln (r_i/r_f) ~,
\eea
where $r_i>r>r_f$ is the valid region for the behavior
(\ref{Asympr3}), $r_i/r_f \sim N_S^{1/2}/n^{1/2}$. 
To have $N_e$ e-folds of inflation, we need $\high{ 
m_S \gtrsim \frac{N_e \mpl}{\sqrt{N_S n}} }$. This may rely on moduli
potentials.
(It is easy to
check that the constant vacuum energy
provided by antibranes sitting in the IR
end is negligible to the inflation.) The Hubble constant is
time-dependent in this case,
$H \approx p/t$, and the inflation is in power law, $a(t) \propto
t^p$. It may be interesting to apply the rescaling to the region 
$r<r_f$.

\section{A multi-throat slow-roll model}
\label{SecMultiSR}
In this appendix, we study the case $\beta \lesssim 1$ in a repulsive
B-throat. We show that
it is generally a combination of the DBI and slow-roll
inflation.
The resulting multi-throat slow-roll model is studied. This Appendix
has some
overlap with an independent paper recently
appeared\cite{Firouzjahi:2005dh}.

We start with the slow-roll case. The scalar velocity is determined by
the non-relativistic equation of motion (\ref{Nonrel}) by neglecting
the first term,
\bea
\dot r \approx - \frac{V'}{3H} ~.
\label{SRspeed}
\eea
This procedure is valid only when the slow-roll condition is
satisfied, $\beta \ll 3$.
This velocity reaches the speed-limit $h^2$ at 
\bea
r \approx \beta H R_B^2/3 ~.
\label{rRel}
\eea

Within this slow-roll region the inflationary e-folds is given by
\bea
N_e = -\frac{T_3}{\mpl^2} \int_r^{r_m} \frac{V}{V'} dr 
= 3 \beta^{-1} \ln (r_m/r) ~,
\eea
where $r_m$ denotes the end of the flat potential and we approximate
it as $R_B$, the extension of the throat. Taking the lower limit
(\ref{rRel}), we get the total number of slow-roll inflationary
e-folds
\bea
(N_{tot})_{sr} &\approx& - 3 \beta^{-1} \ln (\beta H R_B)  
\label{Nsr} \\
&\approx& 3 \beta^{-1} \ln \left( \frac{\mpl}{\beta R_B T_3^{1/2}}
\frac{1}{\sqrt{V_i}} \right) \nonumber \\
&\sim& 6 \beta^{-1} |\ln V_i^{1/4} | ~.
\eea
In the last step, we have neglected all terms other than $V_i$ for
simplicity.\footnote{Even if $\beta$ is not very small as long as it
satisfies the slow-roll condition, for example $\beta
\approx 0.3$, a long period of slow-roll inflation can be achieved
because of
the factor $|\ln V_i^{1/4}|$. For example, if $V_i$ is supplied by an
anti-D3-brane in an A-throat, $|\ln V_i^{1/4}| \approx |\ln h_A|$,
which can be $\sim 10$. But such a period of inflation cannot be
responsible for the CMB since the density perturbations that it
generates is not scale invariant, $n_s -1 \sim {\cal O}(\beta)$.}

Within the relativistic region, the inflaton behaves as
(\ref{rAsymp}), but now the condition (\ref{tUpper2})
\bea
r \ll \beta H R_B^2 
\eea
is stronger than
(\ref{tUpper}). This condition just matches (\ref{rRel}). Taking the
strongest lower bound from Sec.~\ref{SecDBI}, the
DBI inflation then happens within
\bea
-\sqrt{N_B} H^{-1} < t < -\beta^{-1} H^{-1}
~~~~
{\rm or} ~~~~ 
H R_B^2/\sqrt{N_B} < r < \beta H R_B^2
~. 
\eea
The total DBI inflationary e-folds is $\sqrt{N_B}$.

We can state the overall results by varying the parameter $\beta$.
For $1\lesssim \beta \ll \sqrt{N_B}$, we only have the DBI
inflation
and it lasts for $\sqrt{N_B}/\beta$ e-folds. For $0.1 \lesssim \beta
\lesssim 1$, the DBI inflation still proceeds, but its end
starts to deform to
slow-roll, the observable universe is a mixture of
both. For $0 < \beta \lesssim 0.1$, a total $6 \beta^{-1} |\ln
V_i^{1/4}|$ e-folds of slow-roll inflation is smoothly added to the
end of a
total $\sqrt{N_B}$ e-folds of DBI inflation.

\medskip
Following the above discussions, we can have the following
multi-throat slow-roll model. Consider $n_B$ number of D3-branes
rolling out of a B-throat, this time under a flat potential $\beta
\lesssim 0.1$. (For slow-roll, the other directions have to be all
lifted, so the
branes do not roll towards those directions.) The inflationary energy
is provided by $n_A$
anti-D3-branes in the A-throat.

The density perturbations due to the slow-roll period can be
calculated using the standard formula
\bea
\delta_H &=& - \frac{T_3 V^{3/2}}{5 \sqrt{3} \pi n_B^{1/2} \mpl^3 V'}
\\
&\approx& \frac{\sqrt{3}}{5\pi} 
\frac{V_i^{1/2}}{\beta \sqrt{n_B} \mpl r_m}
e^{\beta N_e/3} ~.
\label{DenPertSR}
\eea
The spectral index is
\bea
n_s - 1 \approx -\frac{2}{3} \beta ~, ~~~~ 
\frac{dn_s}{d\ln k} \approx 0 ~.
\eea
The density perturbations and spectral indices of the slow-roll
inflation and the proceeding DBI inflation should be
smoothly connected to each other in the transition region. This can be
seen by evaluating (\ref{DenPertSR}) at $N_e \sim (N_{tot})_{sr}$ given in
(\ref{Nsr}). In terms
of the time delay (\ref{Deltat}), both of them have the same $\delta
r_*$, while $\dot r_*$ transits through (\ref{rRel}). As $N_e$
increases, the density perturbation changes from $\sim e^{\beta
N_e/3}$ to $\sim N_e^2$, so is growing, and then get suppressed at
(\ref{Nec}).

From Eq.~(\ref{DenPertSR}), we can see that the warp factor of the
A-throat cannot be as small as the RS ratio $\sim e^{-30}$, because
$V_i^{1/2} = 
\sqrt{2 n_A} h_A^2$ while $\delta_H \approx 1.9\times 10^{-5}$.
The rescaling mechanism does not help here either (but can
happen). For example suppose
$V_i$ is dominated by a kink in the bulk moduli potential and the
branes gain too much kinetic energy so rescaling happens in the
A-throat. This introduces a factor 
$\alpha^{1/2} N_A^{1/4} h_A/V_i^{1/4}$ as in (\ref{DenPert}). 
Hence $\delta_H \propto h_A
V_i^{1/4}$, still too small if $h_A \sim e^{-30}$.

So let us consider situations similar to Case C in
Sec.~\ref{SecMulti}. For example, using 
$\beta \approx 0.01$, $r_m \approx R_B$, $N_B 
\approx 10^4$, $N_e \approx 60$ and $n_A \approx n_B$, 
we get
\bea
\delta_H \approx 3.5 \frac{T_3^{1/4}}{\mpl} h_A^2 ~.
\eea
If $T_3/\mpl^4 \sim 10^{-3}$, we have $h_A \sim 5 \times
10^{-3}$. 

The cosmic string tension in the A-throat is 
\bea
G\mu_F &=& \sqrt{\frac{g_s}{32\pi}} \frac{T_3^{1/2} h_A^2}{\mpl^2} 
\label{GmuSR1} \\
&\approx& 0.008~ \delta_H^2 h_A^{-2} \sqrt{g_s} ~.
\label{GmuSR2}
\eea
It can take a wide range of value. The upper bound comes from the
experimental tensor
modes bound on the inflationary energy $V_i T_3 = 2 n_A h_A^4 T_3$ 
and we take
it to be $V_i T_3/\mpl^4 < 1.7\times 10^{-8}$ as in
(\ref{TensorBound}). From (\ref{GmuSR1}),
\bea
G\mu_F < 9 \times 10^{-6} \sqrt{g_s/n_A} ~.
\eea
The lower bound comes by setting $h_A \sim 1$ in (\ref{GmuSR2}),
\bea
G\mu_F \gtrsim 3 \times 10^{-12} \sqrt{g_s} ~.
\eea
This range is within the observational ability. The strings in various
throats with (\ref{hHagedorn}), left over from
the Hagedorn transition from the dS epoch, have tension (\ref{GH}).


\begin{thebibliography}{}

%\cite{Guth:1980zm}
\bibitem{Guth:1980zm}
A.~H.~Guth,
``The Inflationary Universe: A Possible Solution To The Horizon And Flatness
Problems,''
Phys.\ Rev.\ D {\bf 23}, 347 (1981).
%%CITATION = PHRVA,D23,347;%%

%\cite{Linde:1981mu}
\bibitem{Linde:1981mu}
A.~D.~Linde,
``A New Inflationary Universe Scenario: A Possible Solution Of The Horizon,
Flatness, Homogeneity, Isotropy And Primordial Monopole Problems,''
Phys.\ Lett.\ B {\bf 108}, 389 (1982).
%%CITATION = PHLTA,B108,389;%%

%\cite{Albrecht:1982wi}
\bibitem{Albrecht:1982wi}
A.~Albrecht and P.~J.~Steinhardt,
``Cosmology For Grand Unified Theories With Radiatively Induced Symmetry
Breaking,''
Phys.\ Rev.\ Lett.\  {\bf 48}, 1220 (1982).
%%CITATION = PRLTA,48,1220;%%





%\cite{Mukhanov:1981xt}
\bibitem{Mukhanov:1981xt}
V.~F.~Mukhanov and G.~V.~Chibisov,
``Quantum Fluctuation And 'Nonsingular' Universe. (In Russian),''
JETP Lett.\  {\bf 33}, 532 (1981)
[Pisma Zh.\ Eksp.\ Teor.\ Fiz.\  {\bf 33}, 549 (1981)].
%%CITATION = JTPLA,33,532;%%

%\cite{Starobinsky:1982ee}
\bibitem{Starobinsky:1982ee}
A.~A.~Starobinsky,
``Dynamics Of Phase Transition In The New Inflationary Universe Scenario And
Generation Of Perturbations,''
Phys.\ Lett.\ B {\bf 117}, 175 (1982).
%%CITATION = PHLTA,B117,175;%%

%\cite{Hawking:1982cz}
\bibitem{Hawking:1982cz}
S.~W.~Hawking,
``The Development Of Irregularities In A Single Bubble Inflationary
Universe,''
Phys.\ Lett.\ B {\bf 115}, 295 (1982).
%%CITATION = PHLTA,B115,295;%%

%\cite{Guth:1982ec}
\bibitem{Guth:1982ec}
A.~H.~Guth and S.~Y.~Pi,
``Fluctuations In The New Inflationary Universe,''
Phys.\ Rev.\ Lett.\  {\bf 49}, 1110 (1982).
%%CITATION = PRLTA,49,1110;%%

%\cite{Bardeen:1983qw}
\bibitem{Bardeen:1983qw}
J.~M.~Bardeen, P.~J.~Steinhardt and M.~S.~Turner,
``Spontaneous Creation Of Almost Scale - Free Density Perturbations In An
Inflationary Universe,''
Phys.\ Rev.\ D {\bf 28}, 679 (1983).
%%CITATION = PHRVA,D28,679;%%

%\cite{Mukhanov:1982nu}
\bibitem{Mukhanov:1982nu}
V.~F.~Mukhanov and G.~V.~Chibisov,
``The Vacuum Energy And Large Scale Structure Of The Universe,''
Sov.\ Phys.\ JETP {\bf 56}, 258 (1982)
[Zh.\ Eksp.\ Teor.\ Fiz.\  {\bf 83}, 475 (1982)].
%%CITATION = SPHJA,56,258;%%

%\cite{Mukhanov:1990me}
\bibitem{Mukhanov:1990me}
V.~F.~Mukhanov, H.~A.~Feldman and R.~H.~Brandenberger,
``Theory Of Cosmological Perturbations. Part 1. Classical Perturbations. Part
2. Quantum Theory Of Perturbations. Part 3. Extensions,''
Phys.\ Rept.\  {\bf 215}, 203 (1992).
%%CITATION = PRPLC,215,203;%%



%\cite{Randall:1999ee}
\bibitem{Randall:1999ee}
L.~Randall and R.~Sundrum,
``A large mass hierarchy from a small extra dimension,''
Phys.\ Rev.\ Lett.\  {\bf 83}, 3370 (1999)
[arXiv:hep-ph/9905221].
%%CITATION = HEP-PH 9905221;%%


%\cite{Klebanov:2000hb}
\bibitem{Klebanov:2000hb}
I.~R.~Klebanov and M.~J.~Strassler,
``Supergravity and a confining gauge theory: Duality cascades and
chiSB-resolution of naked singularities,''
JHEP {\bf 0008}, 052 (2000)
[arXiv:hep-th/0007191].
%%CITATION = HEP-TH 0007191;%%


%\cite{Giddings:2001yu}
\bibitem{Giddings:2001yu}
S.~B.~Giddings, S.~Kachru and J.~Polchinski,
``Hierarchies from fluxes in string compactifications,''
Phys.\ Rev.\ D {\bf 66}, 106006 (2002)
[arXiv:hep-th/0105097].
%%CITATION = HEP-TH 0105097;%%



%\cite{Kachru:2003aw}
\bibitem{Kachru:2003aw}
S.~Kachru, R.~Kallosh, A.~Linde and S.~P.~Trivedi,
``De Sitter vacua in string theory,''
Phys.\ Rev.\ D {\bf 68}, 046005 (2003)
[arXiv:hep-th/0301240].
%%CITATION = HEP-TH 0301240;%%



%\cite{Silverstein:2004id}
\bibitem{Silverstein:2004id}
E.~Silverstein,
``TASI / PiTP / ISS lectures on moduli and microphysics,''
arXiv:hep-th/0405068.
%%CITATION = HEP-TH 0405068;%%



%\cite{Kachru:2003sx}
\bibitem{Kachru:2003sx}
S.~Kachru, R.~Kallosh, A.~Linde, J.~Maldacena, L.~McAllister and S.~P.~Trivedi,
``Towards inflation in string theory,''
JCAP {\bf 0310}, 013 (2003)
[arXiv:hep-th/0308055].
%%CITATION = HEP-TH 0308055;%%

%\cite{Silverstein:2003hf}
\bibitem{Silverstein:2003hf}
E.~Silverstein and D.~Tong,
``Scalar speed limits and cosmology: Acceleration from D-cceleration,''
Phys.\ Rev.\ D {\bf 70}, 103505 (2004)
[arXiv:hep-th/0310221].
%%CITATION = HEP-TH 0310221;%%

%\cite{Hsu:2003cy}
\bibitem{Hsu:2003cy}
J.~P.~Hsu, R.~Kallosh and S.~Prokushkin,
``On brane inflation with volume stabilization,''
JCAP {\bf 0312}, 009 (2003)
[arXiv:hep-th/0311077].
%%CITATION = HEP-TH 0311077;%%

%\cite{Firouzjahi:2003zy}
\bibitem{Firouzjahi:2003zy}
H.~Firouzjahi and S.~H.~H.~Tye,
``Closer towards inflation in string theory,''
Phys.\ Lett.\ B {\bf 584}, 147 (2004)
[arXiv:hep-th/0312020].
%%CITATION = HEP-TH 0312020;%%

%\cite{Pilo:2004mg}
\bibitem{Pilo:2004mg}
L.~Pilo, A.~Riotto and A.~Zaffaroni,
``Old inflation in string theory,''
JHEP {\bf 0407}, 052 (2004)
[arXiv:hep-th/0401004].
%%CITATION = HEP-TH 0401004;%%

%\cite{Burgess:2004kv}
\bibitem{Burgess:2004kv}
C.~P.~Burgess, J.~M.~Cline, H.~Stoica and F.~Quevedo,
``Inflation in realistic D-brane models,''
JHEP {\bf 0409}, 033 (2004)
[arXiv:hep-th/0403119].
%%CITATION = HEP-TH 0403119;%%

%\cite{DeWolfe:2004qx}
\bibitem{DeWolfe:2004qx}
O.~DeWolfe, S.~Kachru and H.~Verlinde,
``The giant inflaton,''
JHEP {\bf 0405}, 017 (2004)
[arXiv:hep-th/0403123].
%%CITATION = HEP-TH 0403123;%%

%\cite{Iizuka:2004ct}
\bibitem{Iizuka:2004ct}
N.~Iizuka and S.~P.~Trivedi,
``An inflationary model in string theory,''
Phys.\ Rev.\ D {\bf 70}, 043519 (2004)
[arXiv:hep-th/0403203].
%%CITATION = HEP-TH 0403203;%%

%\cite{Alishahiha:2004eh}
\bibitem{Alishahiha:2004eh}
M.~Alishahiha, E.~Silverstein and D.~Tong,
``DBI in the sky,''
Phys.\ Rev.\ D {\bf 70}, 123505 (2004)
[arXiv:hep-th/0404084].
%%CITATION = HEP-TH 0404084;%%

%\cite{Berg:2004ek}
\bibitem{Berg:2004ek}
M.~Berg, M.~Haack and B.~Kors,
``Loop corrections to volume moduli and inflation in string theory,''
Phys.\ Rev.\ D {\bf 71}, 026005 (2005)
[arXiv:hep-th/0404087].
%%CITATION = HEP-TH 0404087;%%

%\cite{Blanco-Pillado:2004ns}
\bibitem{Blanco-Pillado:2004ns}
J.~J.~Blanco-Pillado {\it et al.},
``Racetrack inflation,''
JHEP {\bf 0411}, 063 (2004)
[arXiv:hep-th/0406230].
%%CITATION = HEP-TH 0406230;%%

%\cite{Buchel:2004qg}
\bibitem{Buchel:2004qg}
A.~Buchel and A.~Ghodsi,
``Braneworld inflation,''
Phys.\ Rev.\ D {\bf 70}, 126008 (2004)
[arXiv:hep-th/0404151].
%%CITATION = HEP-TH 0404151;%%

%\cite{Chen:2004gc}
\bibitem{Chen:2004gc}
X.~g.~Chen,
``Multi-throat brane inflation,''
arXiv:hep-th/0408084.
%%CITATION = HEP-TH 0408084;%%

%\cite{Berg:2004sj}
\bibitem{Berg:2004sj}
M.~Berg, M.~Haack and B.~Kors,
``On the moduli dependence of nonperturbative superpotentials in brane
inflation,''
arXiv:hep-th/0409282.
%%CITATION = HEP-TH 0409282;%%

%\cite{Shandera:2004zy}
\bibitem{Shandera:2004zy}
S.~E.~Shandera,
``Slow roll in brane inflation,''
arXiv:hep-th/0412077.
%%CITATION = HEP-TH 0412077;%%

%\cite{Firouzjahi:2005dh}
\bibitem{Firouzjahi:2005dh}
H.~Firouzjahi and S.~H.~Tye,
``Brane Inflation and Cosmic String Tension in Superstring Theory,''
arXiv:hep-th/0501099.
%%CITATION = HEP-TH 0501099;%%


%\cite{Becker:2005sg}
\bibitem{Becker:2005sg}
  K.~Becker, M.~Becker and A.~Krause,
  %``M-theory inflation from multi M5-brane dynamics,''
  Nucl.\ Phys.\ B {\bf 715}, 349 (2005)
  [arXiv:hep-th/0501130].
  %%CITATION = HEP-TH 0501130;%%


%\cite{Dvali:1998pa}
\bibitem{Dvali:1998pa}
G.~R.~Dvali and S.~H.~H.~Tye,
``Brane inflation,''
Phys.\ Lett.\ B {\bf 450}, 72 (1999)
[arXiv:hep-ph/9812483].
%%CITATION = HEP-PH 9812483;%%


%\cite{Burgess:2001fx}
\bibitem{Burgess:2001fx}
C.~P.~Burgess, M.~Majumdar, D.~Nolte, F.~Quevedo, G.~Rajesh and R.~J.~Zhang,
``The inflationary brane-antibrane universe,''
JHEP {\bf 0107}, 047 (2001)
[arXiv:hep-th/0105204].
%%CITATION = HEP-TH 0105204;%%

%\cite{Dvali:2001fw}
\bibitem{Dvali:2001fw}
G.~R.~Dvali, Q.~Shafi and S.~Solganik,
``D-brane inflation,''
arXiv:hep-th/0105203.
%%CITATION = HEP-TH 0105203;%%

%\cite{Alexander:2001ks}
\bibitem{Alexander:2001ks}
  S.~H.~S.~Alexander,
  %``Inflation from D - anti-D brane annihilation,''
  Phys.\ Rev.\ D {\bf 65}, 023507 (2002)
  [arXiv:hep-th/0105032].
  %%CITATION = HEP-TH 0105032;%%

%\cite{Shiu:2001sy}
\bibitem{Shiu:2001sy}
  G.~Shiu and S.~H.~H.~Tye,
  %``Some aspects of brane inflation,''
  Phys.\ Lett.\ B {\bf 516}, 421 (2001)
  [arXiv:hep-th/0106274].
  %%CITATION = HEP-TH 0106274;%%

%\cite{Quevedo:2002xw}
\bibitem{Quevedo:2002xw}
  F.~Quevedo,
  %``Lectures on string / brane cosmology,''
  Class.\ Quant.\ Grav.\  {\bf 19}, 5721 (2002)
  [arXiv:hep-th/0210292].
  %%CITATION = HEP-TH 0210292;%%


%\cite{Kachru:2002gs}
\bibitem{Kachru:2002gs}
S.~Kachru, J.~Pearson and H.~Verlinde,
``Brane/flux annihilation and the string dual of a non-supersymmetric  field
theory,''
JHEP {\bf 0206}, 021 (2002)
[arXiv:hep-th/0112197].
%%CITATION = HEP-TH 0112197;%%


%\cite{Chen:2004hu}
\bibitem{Chen:2004hu}
X.~Chen,
``Cosmological rescaling through warped space,''
arXiv:hep-th/0406198.
%%CITATION = HEP-TH 0406198;%%



%\cite{Jones:2002cv}
\bibitem{Jones:2002cv}
N.~Jones, H.~Stoica and S.~H.~H.~Tye,
``Brane interaction as the origin of inflation,''
JHEP {\bf 0207}, 051 (2002)
[arXiv:hep-th/0203163].
%%CITATION = HEP-TH 0203163;%%

%\cite{Sarangi:2002yt}
\bibitem{Sarangi:2002yt}
S.~Sarangi and S.~H.~H.~Tye,
``Cosmic string production towards the end of brane inflation,''
Phys.\ Lett.\ B {\bf 536}, 185 (2002)
[arXiv:hep-th/0204074].
%%CITATION = HEP-TH 0204074;%%

%\cite{Polchinski:2004ia}
\bibitem{Polchinski:2004ia}
J.~Polchinski,
``Introduction to cosmic F- and D-strings,''
arXiv:hep-th/0412244.
%%CITATION = HEP-TH 0412244;%%


%\cite{Jones:2003da}
\bibitem{Jones:2003da}
N.~T.~Jones, H.~Stoica and S.~H.~H.~Tye,
``The production, spectrum and evolution of cosmic strings in brane
inflation,''
Phys.\ Lett.\ B {\bf 563}, 6 (2003)
[arXiv:hep-th/0303269].
%%CITATION = HEP-TH 0303269;%%

%\cite{Pogosian:2003mz}
\bibitem{Pogosian:2003mz}
L.~Pogosian, S.~H.~H.~Tye, I.~Wasserman and M.~Wyman,
``Observational constraints on cosmic string production during brane
inflation,''
Phys.\ Rev.\ D {\bf 68}, 023506 (2003)
[arXiv:hep-th/0304188].
%%CITATION = HEP-TH 0304188;%%


%\cite{Dvali:2003zj}
\bibitem{Dvali:2003zj}
G.~Dvali and A.~Vilenkin,
``Formation and evolution of cosmic D-strings,''
JCAP {\bf 0403}, 010 (2004)
[arXiv:hep-th/0312007].
%%CITATION = HEP-TH 0312007;%%

%\cite{Copeland:2003bj}
\bibitem{Copeland:2003bj}
E.~J.~Copeland, R.~C.~Myers and J.~Polchinski,
``Cosmic F- and D-strings,''
JHEP {\bf 0406}, 013 (2004)
[arXiv:hep-th/0312067].
%%CITATION = HEP-TH 0312067;%%

%\cite{Leblond:2004uc}
\bibitem{Leblond:2004uc}
L.~Leblond and S.~H.~H.~Tye,
``Stability of D1-strings inside a D3-brane,''
JHEP {\bf 0403}, 055 (2004)
[arXiv:hep-th/0402072].
%%CITATION = HEP-TH 0402072;%%

%\cite{Jackson:2004zg}
\bibitem{Jackson:2004zg}
M.~G.~Jackson, N.~T.~Jones and J.~Polchinski,
``Collisions of cosmic F- and D-strings,''
arXiv:hep-th/0405229.
%%CITATION = HEP-TH 0405229;%%

%\cite{Kibble:2004hq}
\bibitem{Kibble:2004hq}
T.~W.~B.~Kibble,
``Cosmic strings reborn?,''
arXiv:astro-ph/0410073.
%%CITATION = ASTRO-PH 0410073;%%

%\cite{Damour:2004kw}
\bibitem{Damour:2004kw}
T.~Damour and A.~Vilenkin,
``Gravitational radiation from cosmic (super)strings: Bursts, stochastic
background, and observational windows,''
arXiv:hep-th/0410222.
%%CITATION = HEP-TH 0410222;%%



%\cite{Burgess:2003ic}
\bibitem{Burgess:2003ic}
C.~P.~Burgess, R.~Kallosh and F.~Quevedo,
``de Sitter string vacua from supersymmetric D-terms,''
JHEP {\bf 0310}, 056 (2003)
[arXiv:hep-th/0309187].
%%CITATION = HEP-TH 0309187;%%

%\cite{Saltman:2004sn}
\bibitem{Saltman:2004sn}
A.~Saltman and E.~Silverstein,
``The scaling of the no-scale potential and de Sitter model building,''
JHEP {\bf 0411}, 066 (2004)
[arXiv:hep-th/0402135].
%%CITATION = HEP-TH 0402135;%%



%\cite{Garriga:1999vw}
\bibitem{Garriga:1999vw}
J.~Garriga and V.~F.~Mukhanov,
``Perturbations in k-inflation,''
Phys.\ Lett.\ B {\bf 458}, 219 (1999)
[arXiv:hep-th/9904176].
%%CITATION = HEP-TH 9904176;%%

%\cite{Armendariz-Picon:1999rj}
\bibitem{Armendariz-Picon:1999rj}
C.~Armendariz-Picon, T.~Damour and V.~Mukhanov,
``k-inflation,''
Phys.\ Lett.\ B {\bf 458}, 209 (1999)
[arXiv:hep-th/9904075].
%%CITATION = HEP-TH 9904075;%%


%\cite{Peebles:1994xt}
\bibitem{Peebles:1994xt}
P.~J.~E.~Peebles,
``Principles of physical cosmology,'' Princeton, USA: Univ. Pr. (1993).
%\href{http://www.slac.stanford.edu/spires/find/hep/www?irn=2994666}{SPIRES
%entry}


%\cite{Liddle:2000cg}
\bibitem{Liddle:2000cg}
A.~R.~Liddle and D.~H.~Lyth,
``Cosmological inflation and large-scale structure,'' Cambridge, UK: Univ. Pr. (2000).
%\href{http://www.slac.stanford.edu/spires/find/hep/www?irn=4458796}{SPIRES entry}



%\cite{Schalm:2004xg}
\bibitem{Schalm:2004xg}
K.~Schalm, G.~Shiu and J.~P.~van der Schaar,
``The cosmological vacuum ambiguity, effective actions, and transplanckian
effects in inflation,''
arXiv:hep-th/0412288.
%%CITATION = HEP-TH 0412288;%%


%\cite{Aharony:1999ti}
\bibitem{Aharony:1999ti}
O.~Aharony, S.~S.~Gubser, J.~M.~Maldacena, H.~Ooguri and Y.~Oz,
``Large N field theories, string theory and gravity,''
Phys.\ Rept.\  {\bf 323}, 183 (2000)
[arXiv:hep-th/9905111].
%%CITATION = HEP-TH 9905111;%%


%\cite{Sen:2002nu}
\bibitem{Sen:2002nu}
A.~Sen,
``Rolling tachyon,''
JHEP {\bf 0204}, 048 (2002)
[arXiv:hep-th/0203211].
%%CITATION = HEP-TH 0203211;%%

%\cite{Sen:2002in}
\bibitem{Sen:2002in}
A.~Sen,
``Tachyon matter,''
JHEP {\bf 0207}, 065 (2002)
[arXiv:hep-th/0203265].
%%CITATION = HEP-TH 0203265;%%

%\cite{Sen:2004nf}
\bibitem{Sen:2004nf}
A.~Sen,
``Tachyon dynamics in open string theory,''
arXiv:hep-th/0410103.
%%CITATION = HEP-TH 0410103;%%

%\cite{Lambert:2003zr}
\bibitem{Lambert:2003zr}
N.~Lambert, H.~Liu and J.~Maldacena,
``Closed strings from decaying D-branes,''
arXiv:hep-th/0303139.
%%CITATION = HEP-TH 0303139;%%


%\cite{Chen:2003xq}
\bibitem{Chen:2003xq}
X.~g.~Chen,
``One loop evolution in rolling tachyon,''
Phys.\ Rev.\ D {\bf 70}, 086001 (2004)
[arXiv:hep-th/0311179].
%%CITATION = HEP-TH 0311179;%%

%\cite{Okuda:2002yd}
\bibitem{Okuda:2002yd}
T.~Okuda and S.~Sugimoto,
``Coupling of rolling tachyon to closed strings,''
Nucl.\ Phys.\ B {\bf 647}, 101 (2002)
[arXiv:hep-th/0208196].
%%CITATION = HEP-TH 0208196;%%


%\cite{Barnaby:2004gg}
\bibitem{Barnaby:2004gg}
N.~Barnaby, C.~P.~Burgess and J.~M.~Cline,
``Warped reheating in brane-antibrane inflation,''
arXiv:hep-th/0412040.
%%CITATION = HEP-TH 0412040;%%

%\cite{McAllister:2004gd}
\bibitem{McAllister:2004gd}
L.~McAllister and I.~Mitra,
``Relativistic D-brane scattering is extremely inelastic,''
arXiv:hep-th/0408085.
%%CITATION = HEP-TH 0408085;%%


%\cite{Peiris:2003ff}
\bibitem{Peiris:2003ff}
H.~V.~Peiris {\it et al.},
``First year Wilkinson Microwave Anisotropy Probe (WMAP) observations:
Implications for inflation,''
Astrophys.\ J.\ Suppl.\  {\bf 148}, 213 (2003)
[arXiv:astro-ph/0302225].
%%CITATION = ASTRO-PH 0302225;%%

%\cite{Tegmark:2003ud}
\bibitem{Tegmark:2003ud}
M.~Tegmark {\it et al.}  [SDSS Collaboration],
``Cosmological parameters from SDSS and WMAP,''
Phys.\ Rev.\ D {\bf 69}, 103501 (2004)
[arXiv:astro-ph/0310723].
%%CITATION = ASTRO-PH 0310723;%%



%\cite{Klemm:1996ts}
\bibitem{Klemm:1996ts}
A.~Klemm, B.~Lian, S.~S.~Roan and S.~T.~Yau,
``Calabi-Yau fourfolds for M- and F-theory compactifications,''
Nucl.\ Phys.\ B {\bf 518}, 515 (1998)
[arXiv:hep-th/9701023].
%%CITATION = HEP-TH 9701023;%%

%\cite{Douglas:2004zg}
\bibitem{Douglas:2004zg}
M.~R.~Douglas,
``Basic results in vacuum statistics,''
Comptes Rendus Physique {\bf 5}, 965 (2004)
[arXiv:hep-th/0409207].
%%CITATION = HEP-TH 0409207;%%


%\cite{Englert:1988th}
\bibitem{Englert:1988th}
F.~Englert, J.~Orloff and T.~Piran,
``Fundamental Strings And Large Scale Structure Formation,''
Phys.\ Lett.\ B {\bf 212}, 423 (1988).
%%CITATION = PHLTA,B212,423;%%



\end{thebibliography}
\end{document}